\documentclass[twocolumn,trackchanges]{aastex61}

\newcommand{\rhoi}{\rho_{\mathrm{i}}}
\newcommand{\rhoh}{\rho_{\mathrm{H}}}
\newcommand{\rhohe}{\rho_{\mathrm{He\,I}}}

\newcommand{\aih}{\alpha_{\rm i\,H}}
\newcommand{\aihe}{\alpha_{\rm i\,He\,I}}
\newcommand{\ahhe}{\alpha_{\rm H\,He\,I}}

\newcommand{\ahi}{\alpha_{\rm H\,i}}
\newcommand{\ahei}{\alpha_{\rm He\,I\,i}}
\newcommand{\aheh}{\alpha_{\rm He\,I\,H}}

\newcommand{\nuih}{\nu_{\rm i\,H}}
\newcommand{\nuihe}{\nu_{\rm i\,He}}

\newcommand{\nuhi}{\nu_{\rm H\,i}}
\newcommand{\nuhei}{\nu_{\rm He\,I\,i}}

\newcommand{\nuhhe}{\nu_{\rm H\,He\,I}}
\newcommand{\nuheh}{\nu_{\rm He\,I\,H}}

\newcommand{\nui}{\nu_{\rm i}}
\newcommand{\nuh}{\nu_{\rm H}}
\newcommand{\nuhe}{\nu_{\rm He\,I}}

\begin{document}

\title{Energy transport and  heating by torsional Alfv\'en waves\\ propagating from the photosphere to the corona in the quiet Sun}

	\shorttitle{Torsional Alfv\'en waves}

\author{Roberto Soler}
\affil{Departament de F\'isica, Universitat de les Illes Balears, E-07122 Palma de Mallorca, Spain}
\affil{Institut d'Aplicacions Computacionals de Codi Comunitari (IAC $^3$), Universitat de les Illes Balears, E-07122 Palma de Mallorca, Spain}

\author{Jaume Terradas}
\affil{Departament de F\'isica, Universitat de les Illes Balears, E-07122 Palma de Mallorca, Spain}
\affil{Institut d'Aplicacions Computacionals de Codi Comunitari (IAC $^3$), Universitat de les Illes Balears, E-07122 Palma de Mallorca, Spain}
\author{Ram\'on Oliver}
\affil{Departament de F\'isica, Universitat de les Illes Balears, E-07122 Palma de Mallorca, Spain}
\affil{Institut d'Aplicacions Computacionals de Codi Comunitari (IAC $^3$), Universitat de les Illes Balears, E-07122 Palma de Mallorca, Spain}
\author{Jos\'e Luis Ballester}
\affil{Departament de F\'isica, Universitat de les Illes Balears, E-07122 Palma de Mallorca, Spain}
\affil{Institut d'Aplicacions Computacionals de Codi Comunitari (IAC $^3$), Universitat de les Illes Balears, E-07122 Palma de Mallorca, Spain}
 
 \email{roberto.soler@uib.es}

\begin{abstract}

In the solar atmosphere, Alfv\'en waves are believed to play an important role in the transfer of energy from the photosphere to the corona and solar wind, and in the heating of the chromosphere. We perform numerical computations to investigate energy transport and dissipation associated with torsional Alfv\'en waves propagating in magnetic flux tubes that expand from the photosphere to the corona in quiet-Sun conditions.  We place a broadband driver at the photosphere that injects a wave energy flux of $10^7$~erg~cm$^{-2}$~s$^{-1}$ and consider Ohm's magnetic diffusion and ion-neutral collisions as dissipation mechanisms. We find that only a small fraction of the driven flux, $\sim 10^5$~erg~cm$^{-2}$~s$^{-1}$, is able to reach coronal heights, but it may be sufficient to partly compensate the total coronal energy loss. The frequency of maximal transmittance is $\sim 5$~mHz for a photospheric field strength of 1~kG and is shifted to smaller/larger frequencies for weaker/stronger fields. Lower frequencies are reflected at the transition region, while higher frequencies are dissipated producing enough heat to balance chromospheric radiative losses. Heating in the low and middle chromosphere is due to Ohmic dissipation, while ion-neutral friction dominates in the high chromosphere. Ohmic diffusion is enhanced by phase mixing because of the expansion of the magnetic field. This effect has the important consequence of increasing the chromospheric dissipation and, therefore, reducing the energy flux that reaches the corona. We provide empirical fits of the transmission coefficient that could be used as input for coronal models.

\end{abstract}


\section{Introduction}

Recent high-resolution observations have shown that Alfv\'enic waves. i.e., incompressible or nearly incompressible magnetohydrodynamic (MHD) waves, are ubiquitous in the solar atmosphere \citep[see, e.g.,][to name a few recent observations]{depontieu2007,jess2009,mcintosh2011,depontieu2014,mortion2015,Srivastava2017,Jafarzadeh2017}. The overwhelming  presence of the waves, together with estimations of their energy, strongly suggest that they may play an important role in the energy balance of the plasma and the propagation of energy through the  atmospheric layers \citep[see, e.g.,][]{hollweg1978,cranmer2005,cargill2011,mathioudakis2013,jess2015}. However, despite the observational evidence, there are still several open issues regarding, for instance, the efficiency of the wave dissipation as a plasma heating mechanism and the ability of the waves to supply a significant amount of energy to the  corona and extended atmosphere that may compensate the continuous energy loss. The present paper aims to shed some light on both relevant issues from a  theoretical point of view.

In the lower atmosphere of the quiet Sun, most of the magnetic flux is concentrated in the network, which is the source of the magnetic field that extends from the photosphere up to the corona. The photospheric network flux is in the form of magnetic  tubes that occupy, typically, 1\%  of the volume in the photosphere and have field strengths of the order of 1~kG \citep{stenflo2000,solanki2000}. It is well-known that magnetic flux tubes  act as waveguides for MHD waves and support a number of different MHD wave modes \citep[see, e.g., a recent summary in][]{jess2015}. In such waveguides, pure Alfv\'en waves take the form of torsional waves, whose restoring force is magnetic tension. By pure Alfv\'en waves we mean Alfv\'en waves that are not coupled with another kind of mode, as it is known that, in general, MHD waves in flux tubes have mixed properties  \citep[see, e.g.,][]{goossens2012}. Torsional Alfv\'en waves produce axisymmetric velocity and magnetic field perturbations that, in a cylindrical tube, are polarized in the azimuthal direction \citep[see, e.g.,][]{erdelyi2007}. In the solar photosphere, observations and numerical simulations show that  horizontal flows \citep[e.g.,][]{Spruit1981,Choudhuri1993,Huang1995,Stangalini2014} and vortex motions \citep[e.g.,][]{Shelyag2011,Shelyag2012,Wedemeyer2012,morton2013} can efficiently drive this kind of incompressible waves in the flux tubes anchored there. In this scenario, part of the mechanical energy of the bulk photospheric motions is converted into wave energy that subsequently propagates  to the upper layers along the magnetic field lines. The estimated driven energy flux, averaged over the whole photosphere, could be as large as $\sim 10^7$~erg~cm$^{-2}$~s$^{-1}$. Thus, in theory, the waves may supply a significant amount of energy to the overlying atmosphere, where part of it could be thermalized.

The accurate theoretical description of the role of the waves in the energy transport in the solar atmosphere requires the use of realistic models. An important ingredient is the expansion with height of the magnetic field. The decrease of the gas pressure with height in the atmosphere results in the radial expansion of the flux tubes. In the lower chromosphere, at a height  of 500-1,000~km above the photosphere, the magnetic field has expanded so much that neighboring flux tubes meet and occupy the whole volume in the upper chromosphere and corona \citep{spruit2000}. Both gravitational stratification of the plasma and expansion of the magnetic field have an important impact on the reflection and transmission properties of the waves as they propagate from the photosphere to the corona. So, the net wave energy flux that can reach the upper layers depends on the amount of reflection in the lower atmosphere \citep[see, e.g.,][]{hollweg1978,hollweg1981,leroy1980,hollweg1982,similon1992,cranmer2005}.

Another important ingredient is the consideration of the  mechanisms that could efficiently dissipate the wave energy. In this regard, partial ionization effects in the chromosphere are essential to correctly describe the chromospheric physics and dynamics \citep[see, e.g.,][]{sykora2012,sykora2017}. It has been shown that partial ionization heavily influences the properties of Alfv\'en waves \citep[see, e.g.,][among others]{piddington1956,osterbrock1961,haerendel1992,khodachenko2004,zaqarashvili2011,soler2013}. In the chromosphere, ion-neutral collisions can efficiently damp Alfv\'en waves and dissipate their energy into the plasma \citep[see, e.g.,][]{leake2005,goodman2011,tu2013,arber2016,shelyag2016}. In addition, electron-neutral collisions enhance the effect of magnetic diffusion, so that electric currents are more efficiently dissipated in partially ionized plasmas than in fully ionized plasmas \citep[see, e.g.,][]{khomenko2012}. Thus, the presence of neutrals is essential to correctly describe the dissipation of waves in the chromosphere.

Here, we aim to study the propagation and dissipation of Alfv\'en wave energy in the lower solar atmosphere. The present paper follows and improves the previous work by \citet{soler2017}. Two important improvements are here incorporated. On the one hand, we abandon the thin flux tube approximation used in the 1.5D model of \citet{soler2017}. Instead, we consider an expanding magnetic flux tube of finite width in which the radial dependence of the axisymmetric wave perturbations is explicitly solved, so  the model is 2.5D. This makes it possible to add a presumably important effect missing from the previous 1.5D models, namely the phase mixing of Alfv\'en waves. On the other hand, besides ion-neutral collisions, here we also consider Ohm's magnetic diffusion as  a dissipation mechanism for the waves, which was not included in  \citet{soler2017}. This addition allows us to perform a better description of wave dissipation in the low and mid chromosphere.

This paper is organized as follows. Section~\ref{sec:model} contains the description of the background atmospheric and magnetic field models. Section~\ref{sec:equations} includes the basic equations and the mathematical expressions used to investigate torsional Alfv\'en waves, while Section~\ref{sec:method} explains the numerical method we have followed to solve those equations. Then, we present and analyze the results in Section~\ref{sec:results}. Subsequently, in Section~\ref{sec:fit} we use the numerical results to provide some empirical fits of the wave energy transmission coefficient, which could be used in future models. Finally, Section~\ref{sec:conc} summarizes our main findings, discusses limitations, and explores some ideas for forthcoming works.

\section{Background atmosphere and magnetic field}
\label{sec:model}

\subsection{Quiet-Sun atmospheric model}

The considered model for the lower solar atmosphere is an improved version of that used in \citet{soler2017}. Here we give a brief summary of how the model is built and refer to \citet{soler2017} for more details.

We use a static, gravitationally stratified  background plasma based on the semi-empirical  quiet-Sun chromospheric model C  of \citet{FAL93}, hereafter FAL93-C, that has been extended to incorporate the low part of the corona. The model provides the variation of the  physical parameters with height, but is invariant in the horizontal direction. The coordinate $z$ represents the vertical coordinate, with $z=0$ corresponding to the top of the photosphere. Hence, the model extends from the base of the photosphere (down to $z= z_{\rm ph} = -100$~km), through the chromosphere, the transition region  (around  $z\approx$~2,200~km), and the low corona (up to $z= z_{\rm c} =$~4,000~km).

\begin{figure}

\plotone{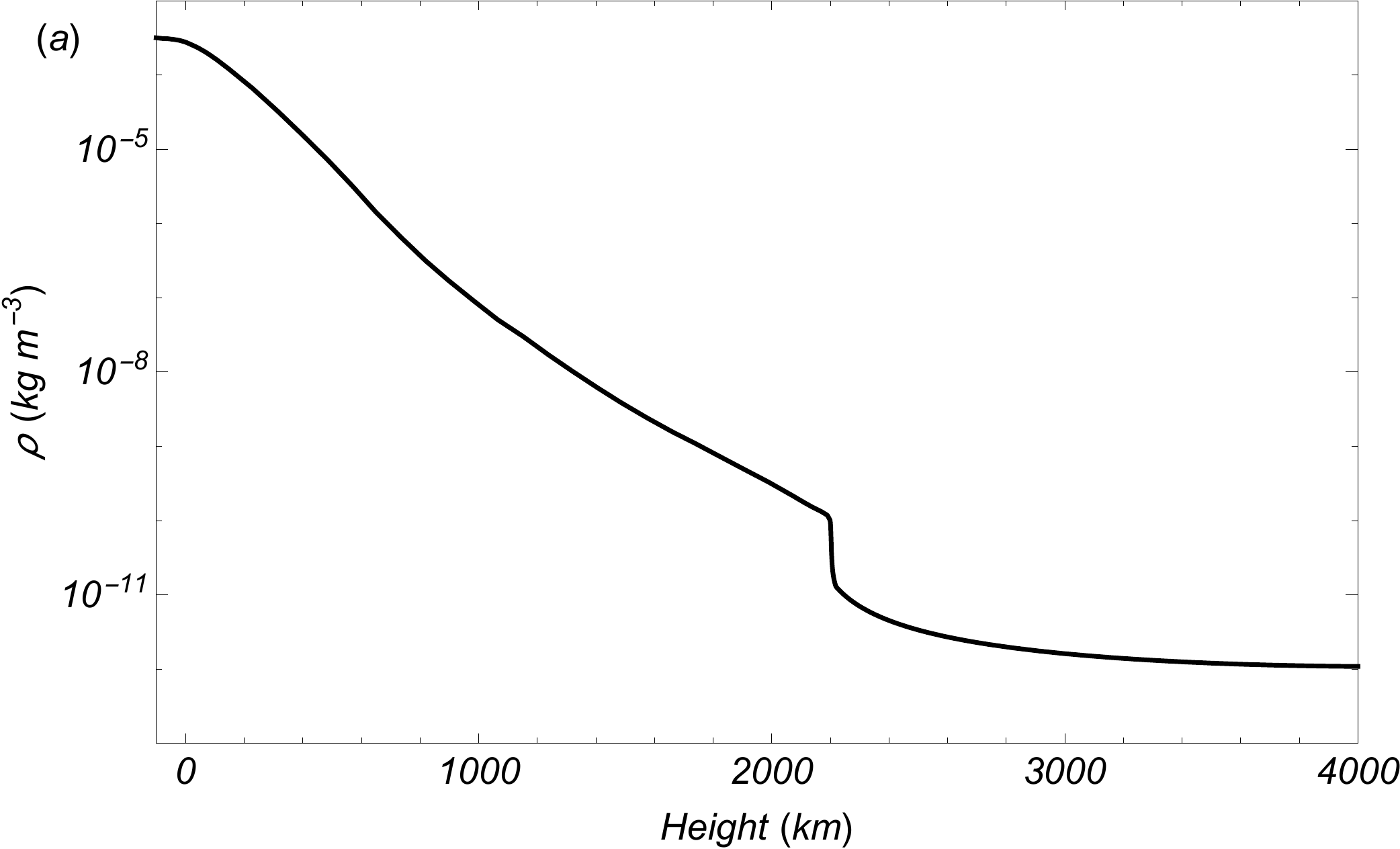}

\plotone{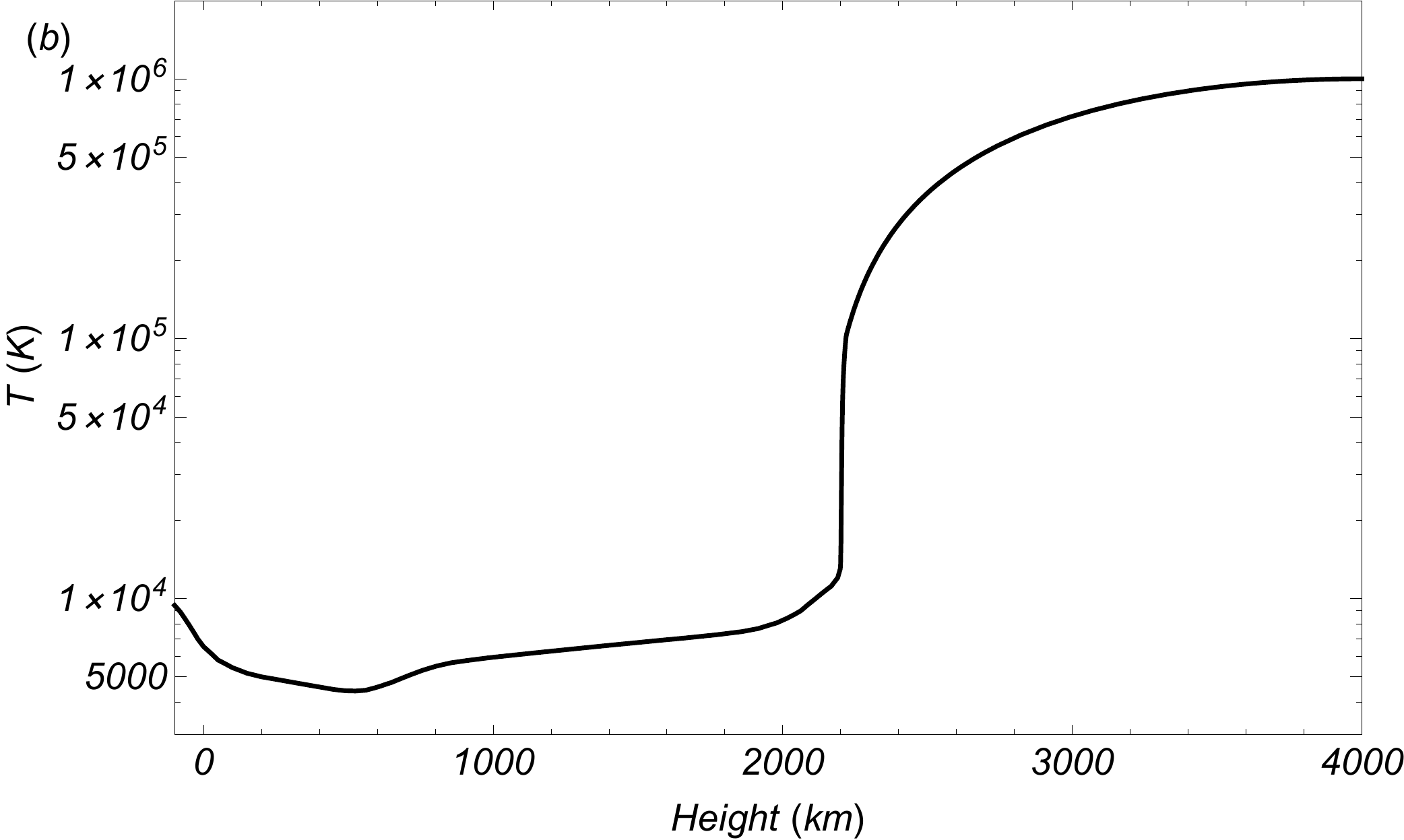}

\plotone{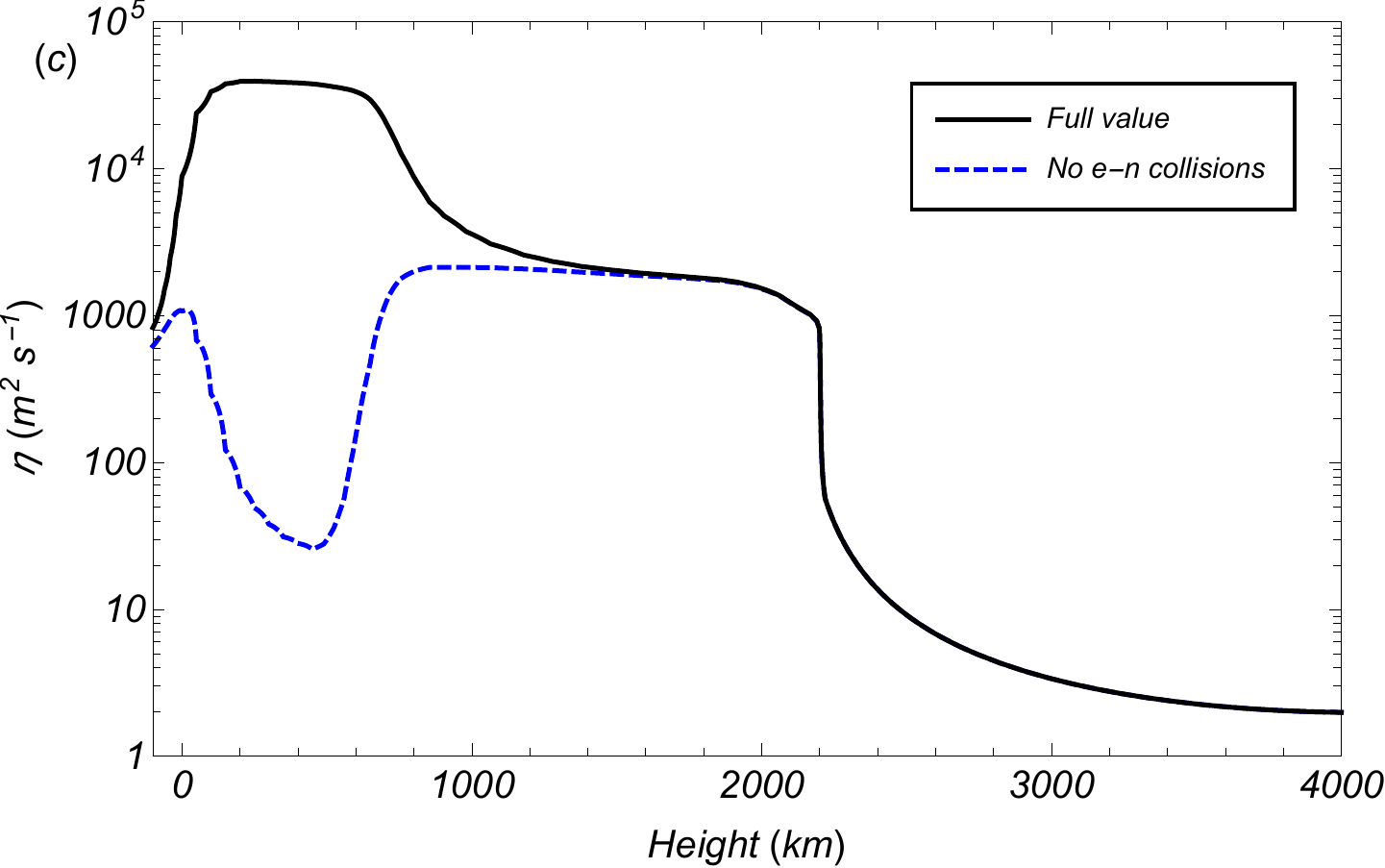}
\caption{Background atmospheric model. Dependence with height above the photosphere of (a) total density, (b) temperature, and (c) Ohmic diffusion  coefficient according to the  chromospheric model C of \citet{FAL93} that has been extended up to 4,000~km above the photosphere to incorporate the low corona.  
\label{fig:background}}
\end{figure}

The plasma  is  composed of hydrogen and helium and is partially ionized. The presence of  species heavier than helium is ignored for the sake of simplicity and because of their negligible abundance. Thus, the considered species are electrons (e), protons (p), neutral hydrogen (H), neutral helium (\ion{He}{1}), singly ionized helium (\ion{He}{2}), and doubly ionized helium (\ion{He}{3}). We denote as $\rho_\beta = n_\beta m_\beta$  the mass density, with $n_\beta$ the number density and $m_\beta$ the particle mass, and as $T_\beta$ the temperature of  a particular species $\beta$, with $\beta=$~e, p, H, \ion{He}{1}, \ion{He}{2}, and \ion{He}{3}. The total mass density of the whole plasma is computed as 
\begin{equation}
\rho = \sum_{\beta} \rho_\beta.
\end{equation}
Since the plasma is highly collisional (see Section~\ref{sec:collisions}), there is a strong thermal coupling between all species. Consequently, we use the same background temperature, $T$, for all species. This temperature is the one provided in the FAL93-C model. In the following formulae, we formally consider different temperatures for each species in order to give the expressions in their most general form.  Figures~\ref{fig:background}(a) and (b)  display the variation with height of the  total density and the temperature according to the FAL93-C model.

\subsection{Collisional effects}
\label{sec:collisions}

The different species that form the plasma are assumed to interact by means of elastic particle  collisions \citep[see, e.g.,][]{brag,schunk1977,draine86}.  Elastic collisions are interactions between two different species in which the total momentum and energy are conserved and the total number of particles of one specific species in a volume element does not change because of the collisions. Collisions of this kind are, e.g., momentum-transfer collisions and charge-exchange collisions. Conversely, in the case of inelastic collisions there are processes that convert particles of one species into another as, e.g., ionization and recombination processes, so that the total number of particles of one specific species is not conserved. The treatment of inelastic collisions is more complicated. Several source and sink terms have to be included in the continuity, momentum, and energy equations of the colliding species and the collision term cannot always be explicitly written. Only elastic collisions are considered here.

We denote by ${\bf R}_{\beta \beta'}$ the exchange of momentum between species $\beta$ and $\beta'$ because of collisions, namely
\begin{equation}
{\bf R}_{\beta \beta'} = \alpha_{\beta \beta'} \left( {\bf v}_\beta -  {\bf v}_{\beta'} \right), \label{eq:erre}
\end{equation} 
where ${\bf v}_\beta$ and ${\bf v}_{\beta'}$ are the  velocities  of species $\beta$ and $\beta'$, so that $\left( {\bf v}_\beta -  {\bf v}_{\beta'} \right)$ is the velocity drift, and $\alpha_{\beta \beta'}$ is the so-called friction or momentum-transfer coefficient. The presence of  velocity drifts between species causes the redistribution of momentum within the plasma. This has the consequence that any perturbation superimposed on the static background equilibrium is damped, although there is no loss of total energy. The energy of the perturbations is simply transformed into internal energy of the plasma. The collisional interaction causes the conversion from kinetic to  internal energy and the transfer of heat between colliding species. We denote by $Q_{\beta \beta'}$ the heat transfer between species $\beta$ and $\beta'$ because of collisions, namely
\begin{eqnarray}
Q_{\beta \beta'} = \frac{2  \alpha_{\beta \beta'}}{m_\beta + m_{\beta'}} &&  \left[\frac{A_{\beta \beta'}}{2} k_{\rm B} \left( T_\beta - T_{\beta'} \right) \right. \nonumber \\ && \left. + \frac{1}{2} m_\beta  \left( {\bf v}_\beta -  {\bf v}_{\beta'} \right)^2 \right], \label{eq:qheat}
\end{eqnarray}
where $k_{\rm B}$ is Boltzmann's constant and $A_{\beta \beta'}$ is a parameter whose value is 4 for electon-neutral collisions and 3 for the other types of collisions \citep[see][]{draine86}. The heat-transfer term, $Q_{\beta \beta'}$, has two contributions. On the one hand, the first term on the right-hand side of Equation~(\ref{eq:qheat}) accounts for the exchange of heat because of the different temperature of the colliding species. This contribution is positive or negative depending on the relative temperature of the two species, meaning that the role of this term is to equalize the temperatures. According to this, in the static background atmosphere the temperature differences between species vanish in very short time scales of the order of the particle collision times \citep[see also][]{spitzer1962}. This justifies the use of a common temperature for all species.  On the other hand, the second term on the right-hand side of Equation~(\ref{eq:qheat}) accounts for the conversion from kinetic energy to internal energy during the collisions. This term is quadratic in the velocity drift, meaning that its contribution is always positive. This term represents a net heating because of the collisional friction.

The friction coefficients,  $\alpha_{\beta \beta'}$, measure the strength of the interaction between species and depend on the local plasma parameters and  the type of collisions. In the case of collisions between two electrically charged species, namely $q$ and $q'$, the interaction is Coulombian and the friction coefficient is computed as \citep[e.g.,][]{spitzer1962,brag}
\begin{equation}
\alpha_{qq'} = \frac{n_q n_{q'} Z_q^2 Z_{q'}^2 e^4 \ln \Lambda_{qq'}}{6\pi\sqrt{2\pi} \epsilon_0^2 m_{q q'} \left( k_{\rm B} T_q /m_q +   k_{\rm B} T_{q'} /m_{q'} \right)^{3/2}}, \label{eq:fric}
\end{equation}
 where  $m_{qq'} = m_q m_{q'}/\left( m_q + m_{q'} \right)$ is the reduced mass, $Z_q$ and $Z_{q'}$ are the signed charged number of the particles, $e$ is the electron charge,  $\epsilon_0$ is the permittivity of free space,  and $\ln \Lambda_{qq'}$ is the so-called Coulomb's logarithm given by \citep[e.g.,][]{spitzer1962,vranjes2013}
\begin{eqnarray}
\ln\Lambda_{q q'} = \ln && \left[ \frac{12  \pi  \epsilon_0^{3/2}  k_{\rm B}^{3/2} \left( T_q + T_{q'} \right)}{\left| Z_q Z_{q'} \right| e^3} \right. \nonumber \\ & & \times \left.\sqrt{\frac{T_q T_{q'}}{Z_q^2 n_q T_{q'}+Z_{q'}^2 n_{q'} T_{q}}}  \right].
\end{eqnarray}
On the other hand, when the collisions involve at least one neutral species, the scattering of particles is produced mainly because of direct impacts. Then, the friction coefficient in the approximation of small velocity drift can be cast as \citep[e.g.,][]{brag,chapman1970,draine86}
\begin{equation}
\alpha_{\beta \beta'} =  n_{\beta} n_{\beta'} m_{\beta \beta'}  \frac{4}{3}\sigma_{\beta\beta'} \sqrt{\frac{8}{\pi} \left(\frac{k_{\rm B}T_\beta}{m_\beta} + \frac{k_{\rm B}T_{\beta'}}{m_{\beta'}} \right)}, \label{eq:fricneu}
\end{equation}
where  $\sigma_{\beta \beta'}$ is the collision cross-section, which can be taken independent of temperature for plasma conditions in the lower solar atmosphere \citep[see, e.g.,][]{dickinson1999,Lewkow2012,vranjes2013}.  The considered cross-sections take into account charge-exchange effects.  We note that, in all cases, the friction coefficients are symmetric if the two species that collide are interchanged. The total friction coefficient of a species $\beta$ with all the other species is
\begin{equation}
\alpha_\beta = \sum_{\beta' \neq \beta} \alpha_{\beta\beta'}.
\end{equation}
In turn, we denote by $\nu_{\beta\beta'}$ the collision frequency of species $\beta$ with species $\beta'$ and is computed from the friction coefficient as
\begin{equation}
\nu_{\beta\beta'} = \frac{\alpha_{\rm \beta\beta'}}{\rho_\beta}.
\end{equation}
Unlike the friction coefficient, the collision frequency has a more obvious physical meaning. The collision frequency $\nu_{\beta\beta'}$ measures, statistically, the number of encounters or interactions of one particle of species $\beta$ with particles of species $\beta'$ per unit time. The collision frequency is generally not symmetric if the colliding species are interchanged, because of the different densities and effective cross-sections that the two colliding species may have. The total collision frequency of species $\beta$ with all the other species is computed as
\begin{equation}
\nu_\beta = \sum_{\beta' \neq \beta} \nu_{\beta\beta'} = \frac{\alpha_\beta}{\rho_\beta}.
\end{equation}

 For the purpose of the investigation of Alfv\'en waves,  all ions, namely protons (p), singly ionized helium  (\ion{He}{2}), and doubly ionized helium (\ion{He}{3}) are assumed to form a single species, which we generally call ``ions''. We shall use the subscript i to refer to this ionic species. We define the total density, $\rhoi$, and center-of-mass velocity, ${\bf v}_{\rm i}$, of the ions as
\begin{eqnarray}
\rhoi &\equiv& \rho_{\rm p} + \rho_{\rm He\,II} + \rho_{\rm He\,III}, \\
{\bf v}_{\rm i} &\equiv& \frac{\rho_{\rm p} {\bf v}_{\rm p} + \rho_{\rm He\,II} {\bf v}_{\rm He\,II} + \rho_{\rm He\,III} {\bf v}_{\rm He\,III}}{\rhoi}.
\end{eqnarray}
Furthermore, we assume that all ions are strongly coupled, so that their individual velocity drifts are much smaller than their center-of-mass velocity. Therefore, we can approximate 
\begin{equation}
{\bf v}_{\rm p} \approx {\bf v}_{\rm He\,II}  \approx {\bf v}_{\rm He\,III} \approx {\bf v}_{\rm i}.
\end{equation}
The assumption of neglecting the ion-ion drifts is justified as long as the frequency of the Alfv\'en waves remains much lower than the cyclotron frequencies of the individual ions \citep[see, e.g.,][]{martinez2016}. The cyclotron frequencies are about $10^6$~rad~s$^{-1}$ in the magnetized lower solar atmosphere. In addition, we define the global friction coefficient of the ions  by adding the individual coefficients as
\begin{equation}
\alpha_{\rm i \beta} \equiv \alpha_{\rm p\beta} + \alpha_{\rm He\,II \beta} + \alpha_{\rm He\,III \beta},
\end{equation}
with $\beta =$~e, H and \ion{He}{1}. In turn, the global collision frequency of the ions with species $\beta$ is simply $\nu_{\rm i\beta} = \alpha_{\rm i \beta} / \rhoi$, so that the total collision frequency of the ions is
\begin{equation}
\nui = \sum_{\beta \neq \rm i} \nu_{\rm i\beta}.    
\end{equation}

Velocity drifts between electrically-charged species induce electric currents, $\bf J$, according to the expression
\begin{equation}
{\bf J} = e \sum_{q} Z_q n_{q} {\bf v}_q, \label{eq:current}
\end{equation}
where $q=$~e, p, \ion{He}{2}, and \ion{He}{3}. Consequently, these currents are diffused by the collisions of electrons with all the other species, i.e., resistivity or magnetic diffusion.  In terms of the electron total friction coefficient, the coefficient of resistivity, $\eta$, is given by \citep[see, e.g.,][]{khomenko2012}
\begin{equation}
\eta = \frac{\alpha_{\rm e}}{\mu e^2 n_{\rm e}^2},
\end{equation}
where $\mu$ is the magnetic permeability. Figure~\ref{fig:background}(c) displays the variation with height of $\eta$ according to the FAL93-C model. In a partially ionized plasma, as the chromosphere, electron-neutral collisions greatly enhance the value of $\eta$. To highlight this effect, we have overplotted in  Figure~\ref{fig:background}(c) the value of $\eta$ obtained when electron-neutral collisions are ignored. We see that neutrals play a predominant role in setting the value of $\eta$ in the low chromosphere.

\subsection{Magnetic flux tube model}

The magnetic field configuration used here is made of a vertical magnetic flux tube that is embedded in the background atmosphere and expands with height over the photosphere. We use cylindrical coordinates, namely $r$, $\varphi$, and $z$, that denote the radial, azimuthal, and vertical coordinates, respectively. We assume that the flux tube is untwisted and azimuthally symmetric, so that the magnetic field is invariant in $\varphi$ and there is no component in that direction.  Thus, the magnetic field is expressed as 
\begin{equation}
{\bf B} = B_r \left( r, z \right) \hat{e}_r + B_z \left( r, z \right) \hat{e}_z,
\end{equation}
where $B_r \left( r, z \right)$ and $B_z \left( r, z \right)$ are the radial and longitudinal (vertical) components. We assume that this flux tube is in equilibrium, so that the background magnetic field does not evolve with time. Unlike in \citet{soler2017}, we do not restrict ourselves to the thin flux tube approximation, and so allow the magnetic tube to have an arbitrary expansion factor.

A non-potential magnetic field generates electric currents through Ampere's Law, namely
\begin{equation}
{\bf J} = \frac{1}{\mu} \nabla \times {\bf B}.
\end{equation}
In turn, these currents induce velocity drifts between electrically-charged species according to Equation~(\ref{eq:current}), so that the magnetic field is eventually diffused by resistivity because of electron collisions. The diffusion of non-potential magnetic fields in the partially ionized chromosphere has been explored by, e.g., \citet{arber2009,khomenko2012,shelyag2016}. In our  model the consideration of a non-potential background magnetic field is incompatible with the assumption that the background plasma is static and that the magnetic tube itself is in equilibrium. Therefore, in order to satisfy both conditions, we consider a potential, current-free magnetic field, so that ${\bf J} = \bf 0$ in the background. We note that a potential magnetic field is also force-free because ${\bf J} \times {\bf B} = \bf 0$. A  potential flux tube was also used by \cite{brady2016} although in Cartesian coordinates. Hence, we express the background magnetic field as
\begin{equation}
{\bf B} = - \nabla \phi \left( r,z \right),
\end{equation}
where $\phi \left( r,z \right)$ is the magnetic scalar potential. The magnetic field components are explicitly related to the potential as
\begin{equation}
B_r \left( r,z \right) = - \frac{\partial \phi \left( r,z \right)}{\partial r}, \qquad B_z \left( r,z \right) = - \frac{\partial \phi \left( r,z \right)}{\partial z}.
\end{equation} 
Because of the divergence-free condition of the magnetic field, $\nabla \cdot {\bf B} = 0$, the magnetic potential satisfies Laplace's equation, namely
\begin{equation}
\nabla^2 \phi \left( r,z \right) = 0. \label{eq:laplace}
\end{equation}

\begin{figure*}
\plottwo{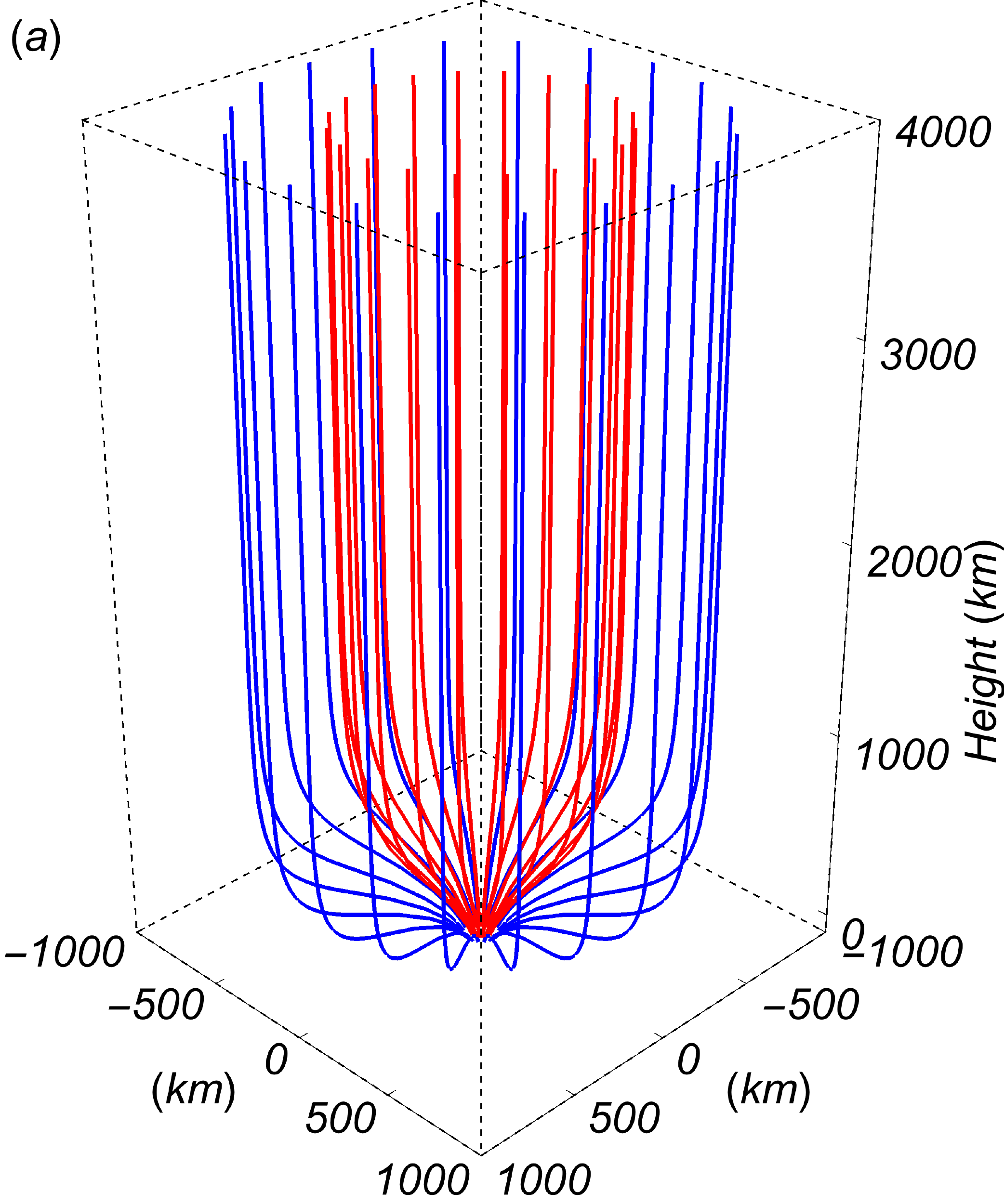}{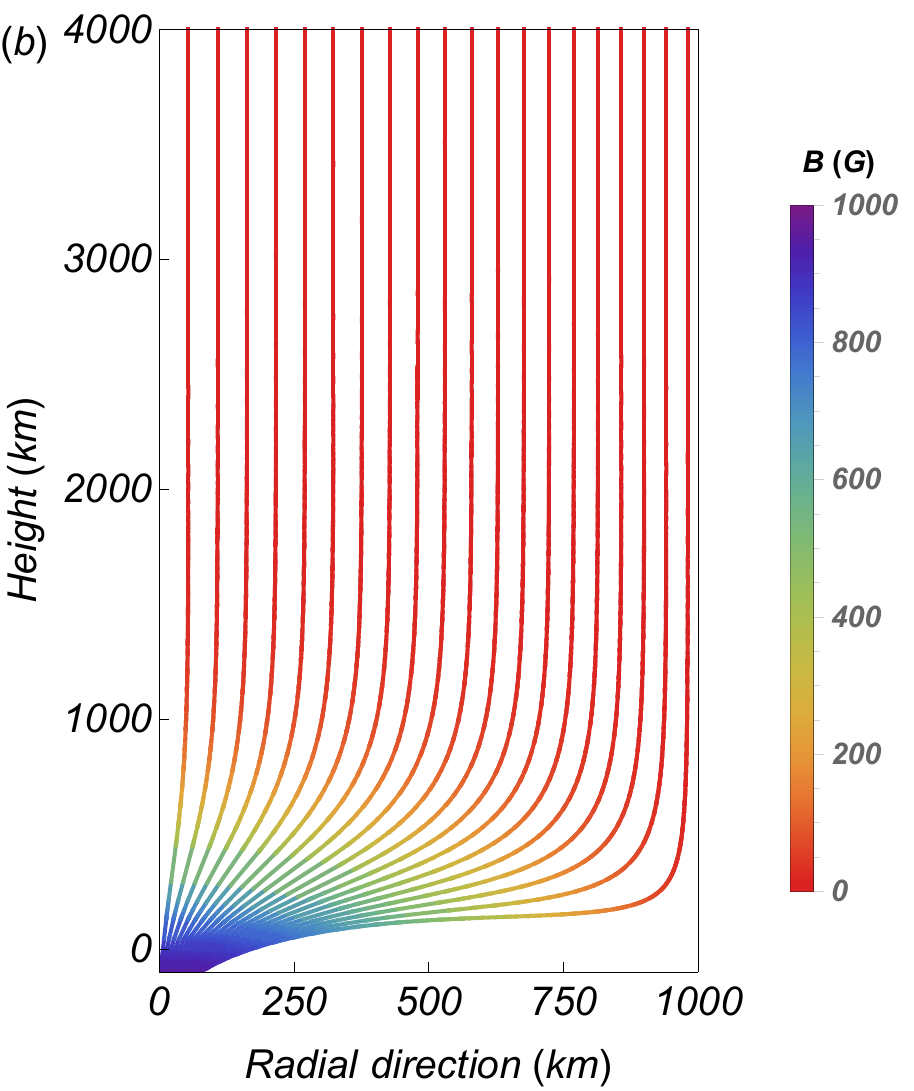}
\caption{(a) Visualization in 3D of the expanding magnetic flux tube  embedded in the stratified lower atmosphere. The red and blue lines outline some selected magnetic field lines that cross the photosphere at $r=0.4 R = 40$~km and $r=0.7 R = 70$~km, respectively. (b) Representation of the magnetic field lines in the $rz$-plane.  The color gradient denotes the variation of the magnetic field strength along the field lines from the photospheric field $B_{\rm ph}$ to the coronal field $B_{\rm c}$.  In this example, we have considered $B_{\rm ph} = 1$~kG and $B_{\rm ph} = 10$~G. For visualization purposes, the horizontal and vertical axes are not to scale.
\label{fig:model}}
\end{figure*}

The potential magnetic flux tube is constructed by numerically solving Equation~(\ref{eq:laplace}) in the numerical domain that extends, vertically, from the base of the photosphere ($z= z_{\rm ph}$) to the low corona ($z=z_{\rm c}$) and, horizontally, from the center of the tube ($r=0$) to a prescribed maximum value of the radial coordinate ($r=r_{\rm max}$). In this context, $r_{\rm max}$ represents the radial distance at which the considered flux tube merges neighboring tubes of the network. In addition, we need to consider appropriate boundary conditions. At the lower photospheric boundary, we follow \cite{brady2016} and prescribe the potential  to represent a flux patch of size $R$, namely
\begin{equation}
\phi = \phi_0 \exp\left( - \frac{r^2}{R^2}\right), \qquad \textrm{at} \qquad z=z_{\rm ph},
\end{equation}
where $R=100$~km is a measure of the flux tube radius at the photosphere and $\phi_0$ is a constant that depends on the value of the magnetic field strength imposed at the center of the photospheric patch, namely $B_{\rm ph}$.  At the upper coronal boundary, we assume a purely vertical and uniform field of strength $B_{\rm c} = 10$~G in all cases, so that the upper boundary condition for the potential is 
\begin{equation}
\frac{\partial \phi}{\partial z} = - B_{\rm c}, \qquad \textrm{at} \qquad z=z_{\rm c}.
\end{equation}
Regarding the boundary conditions at $r=0$ and $r=r_{\rm max}$, we assume $B_r = 0$, so that the condition for the potential is 
\begin{equation}
\frac{\partial \phi }{\partial r} = 0, \qquad \textrm{at} \qquad r=0 \quad \textrm{and} \quad r=r_{\rm max}.
\end{equation}
In all cases we take $r_{\rm max} = 10 R =$~1,000~km.

Figure~\ref{fig:model} displays an example of the potential magnetic flux tube model constructed with the above-mentioned method in the case that the field strength in the photospheric patch is $B_{\rm ph} = 1$~kG. We note that the strong expansion of the field lines occurs at heights below 1,000~km. The field lines are already nearly vertical when the flux tube crosses the transition region around 2,000~km above the photosphere.

\section{Mathematical formalism}
\label{sec:equations}

\subsection{Basic governing equations}
\label{sec:threefluid}

For the purpose of the investigation of Alfv\'en waves, the different species in the plasma are treated as separate fluids that interact by means of particle collisions and electromagnetic fields. All ions are considered together as part of a single ionic fluid (i), while neutral hydrogen (H) and neutral helium (\ion{He}{1}) are treated as two separate neutral fluids with different velocities. 

A special treatment is used in the case of electrons (e).  The inertia of electrons is neglected owing to their small mass, which allows us to obtain the generalized Ohm's law and the corresponding magnetic induction equation from the electron momentum equation. This process can be found in, e.g., \citet{zaqarashvili2011,Khomenko2014,martinez2017}, among others, and is not repeated here for the sake of simplicity. Also because of the negligible electron inertia, we assume that the inertia of the much heavier ions and neutrals is not modified by electron-ion and electron-neutral collisions. However, those collisions have a very important effect on the dynamics of the lighter electrons. Electron collisions cause diffusion of currents, which is mathematically represented by the resistivity or Ohm's diffusion term in the magnetic induction equation \citep[see, e.g.,][]{khomenko2012,Khomenko2014}. 

 Summarizing, the  plasma is composed of three distinct fluids: one ionic fluid and two neutral fluids, while the dynamic of electrons is indirectly included through the magnetic induction equation. Hence, the basic equations in the present three-fluid model  are the momentum, continuity, and energy equations for each fluid, and the magnetic induction equation. The full form of these multi-fluid equations can be checked in, e.g., \citet{zaqarashvili2011,Khomenko2014,martinez2017,ballesterSSR}.

Now, we give the basic equations that are used to study the propagation of Alfv\'en waves. Following \citet{soler2017}, we assume that the Alfv\'en waves produce small perturbations over the static background that are  well described by the linear regime of the equations. Hence, the governing equations are linearized:  each physical variable is expressed as a background value plus a perturbation, and only linear terms in the perturbations are retained. The resulting relevant equations for Alfv\'en waves only involve the  perturbations of the velocity  of the three fluids, ${\bf v'_\beta}$, with $\beta=$~i, H, and \ion{He}{1}, and the  perturbation of the magnetic field, $\bf B'$.  These linearized governing equations for Alfv\'en waves are
\begin{eqnarray}
\rhoi \frac{\partial {\bf v'_{\rm i}}}{\partial t} &=& {\bf J'} \times {\bf B} - \aih \left( {\bf v'_{\rm i}} - {\bf v'_{\rm H}}  \right) \nonumber \\
&& - \aihe \left( {\bf v'_{\rm i}} - {\bf v'_{\rm He\,I}}  \right), \label{eq:momi} \\
\rhoh \frac{\partial {\bf v'_{\rm H}}}{\partial t} &=&  - \ahi \left( {\bf v'_{\rm H}} - {\bf v'_{\rm i}}  \right) \nonumber \\
&&  - \ahhe \left( {\bf v'_{\rm H}} - {\bf v'_{\rm He\,I}}  \right), \label{eq:momh} \\
\rhohe \frac{\partial {\bf v'_{\rm He\,I}}}{\partial t} &=&  - \ahei \left( {\bf v'_{\rm He\,I}} - {\bf v'_{\rm i}}  \right) \nonumber \\
&&  - \aheh \left( {\bf v'_{\rm He\,I}} - {\bf v'_{\rm H}}  \right), \label{eq:momhe} \\
\frac{\partial \bf B'}{\partial t} &=& \nabla \times \left( {\bf v'_{\rm i}} \times {\bf B} \right) - \nabla \times \left( \mu \eta  {\bf J'} \right) \nonumber \\
&&- \nabla \times \left( \frac{1}{e n_{\rm e}} {\bf J'}\times {\bf B} \right), \label{eq:induction}
\end{eqnarray}
where ${\bf J'} = \left( \nabla \times {\bf B'} \right)/\mu$ is the perturbation of the electric current and the rest of quantities have already been defined before.

Equation~(\ref{eq:momi}) is the linearized momentum equation of the ionic fluid and includes  the Lorentz force, ${\bf J'} \times {\bf B}$, and the frictional forces due to collisions as defined in Equation~(\ref{eq:erre}). Equations~(\ref{eq:momh}) and (\ref{eq:momhe}) are  the corresponding momentum equations of neutrals and only include the frictional forces. Gas pressure terms are not present in the linearized momentum equations because Alfv\'en waves are incompressible and do not produce pressure perturbations in the linear regime. In addition, the gravity force is also absent from Equations~(\ref{eq:momi})--(\ref{eq:momhe}) because the magnetic flux tube is vertical and  the Alfv\'en wave motions are confined in horizontal planes perpendicular to the direction of gravity. 

Equation~(\ref{eq:induction}) is the magnetic induction equation and  formally includes Ohm's diffusion (resistivity) and Hall's term, i.e., the second and third terms on the right-hand side, respectively. However, in the present study Hall's term is neglected.  \citet{pandey2008} and \citet{martinez2017}, among others, showed that in a partially ionized plasma Hall's term becomes important for wave frequencies larger than an effective cyclotron frequency called Hall's frequency, $\omega_{\rm H}$, given by
\begin{equation}
\omega_{\rm H} = \frac{ \Omega_{\rm p}}{1 + \rho_{\rm n}/\rhoi},
\end{equation}
were $\rho_{\rm n} = \rhoh + \rhohe$	is the total neutral density and $\Omega_{\rm p} = e \left| {\bf B} \right|/m_{\rm p}$ is the proton cyclotron frequency. In the lower chromosphere, $\Omega_{\rm p} \sim 10^6$~rad~s$^{-1}$ and $\rho_{\rm n}/\rhoi \sim 10^5$, so that $\omega_{\rm H} \sim 10$~rad~s$^{-1}$.  The largest  wave frequency considered in the present work is 300 mHz ($\approx 1.88$~rad~s$^{-1}$), which is well below $\omega_{\rm H}$.  In addition, results of \citet{soler2015overdamped} and \citet{arber2016} suggest that Hall's term has a minor impact on the  damping of Alfv\'en waves and the associated heating rates in the chromosphere.  Hall's term is ignored from here on.

\subsection{Steady state of torsional wave propagation}
\label{sec:torsionalwaves}

We consider Alfv\'en waves of torsional type. Hence, we take $\partial/\partial\varphi = 0$ and assume the velocity and magnetic field perturbations to be strictly polarized in the azimuthal direction, namely
\begin{equation}
{\bf v'_\beta} =  v'_{\beta,\varphi} \hat{e}_\varphi, \qquad {\bf B'} =  B'_\varphi \hat{e}_\varphi.
\end{equation}
In addition, we consider the steady state of wave propagation. Thus, we express the temporal dependence of perturbations as $\exp\left(-i\omega t\right)$, where $\omega$ is the angular frequency of the waves. The prescribed temporal dependence implies that the driver that excites the waves  acts continuously and has been working for a sufficiently long time, so that the waves have had time to propagate and reflect along the whole domain and a stationary pattern has been achieved. The same method was used in, e.g., \citet{goodman2011} and \citet{soler2017}, and it allows us to drop the temporal dependence of the problem while retaining the spatial dependence. At this stage, we do not need to specify the form of the wave driver, although we later assume that it is located at the photosphere.  Detailed information about the specific form of the wave driver used in the computations is given in Section~\ref{sec:bc} when discussing the boundary conditions at the photosphere.

Although the mathematical expressions given below are written using the angular frequency, $\omega$ (given in rad~s$^{-1}$), when discussing the results we shall use the linear frequency, $f$ (given in Hz). Both are related by  $f=\omega/2\pi$, so that the period of the wave  is simply $1/f$.

An alternative approach would be performing time-dependent  simulations, which would provide the full temporal dependence of the waves including any transient \citep[see, e.g.,][]{tu2013,arber2016,brady2016,shelyag2016}. However, multi-fluid time-dependent  simulations are computationally much more expensive than the present steady-state method when very high spatial resolutions are needed in the lower chromosphere \citep[see][]{goodman2011,soler2015scales}. At present, detailed parameter studies, as the ones performed here, are not practical with 2D or 3D time-dependent  simulations. 

Thanks to the temporal dependence $\exp\left(-i\omega t\right)$, we can combine Equations~(\ref{eq:momh}) and (\ref{eq:momhe}) to express the $ v'_{\rm H,\varphi}$ and $v'_{\rm He\,I,\varphi}$ in terms of $v'_{\rm i,\varphi}$ as
\begin{eqnarray}
 v'_{\rm H,\varphi} &=&   \frac{i \nuhi \left( \omega + i\nuhe \right) - \nuhhe \nuhei}{\left( \omega + i\nuh \right)\left( \omega + i\nuhe \right) + \nuhhe\nuheh}   v'_{\rm i,\varphi}, \label{eq:alfv2} \\
 v'_{\rm He\,I,\varphi} &=&  \frac{i \nuhei \left( \omega + i\nuh \right) - \nuheh \nuhi}{\left( \omega + i\nuh \right)\left( \omega + i\nuhe \right) + \nuhhe\nuheh}  v'_{\rm i,\varphi}. \label{eq:alfv3}
\end{eqnarray} 
In addition, from Equation~(\ref{eq:momi}) we can write $v'_{\rm i,\varphi}$ as a function of $B'_\varphi$ as
\begin{equation}
v'_{\rm i,\varphi} = \frac{i}{\omega} \frac{1}{\mu\rho_{\rm eff}} \frac{1}{r} {\bf B} \cdot \nabla \left( r B'_\varphi \right), \label{eq:vi}
\end{equation}
with $\rho_{\rm eff}$  the effective plasma density felt by the Alfv\'en waves. The effective density is a function of the wave frequency and is given by
\begin{equation}
\rho_{\rm eff} = \frac{\Omega_{\rm col}}{\omega} \rhoi, \label{eq:effectiverho}
\end{equation}
where $\Omega_{\rm col}$ is a complex quantity defined in \citet{soler2017} that has dimensions of frequency and contains $\omega$ and all the collision frequencies, namely
\begin{eqnarray}
\Omega_{\rm col} & = & \omega + i \nui +  \frac{\nuih \nuhi \left( \omega + i\nuhe \right) + i \nuih \nuhhe \nuhei}{\left( \omega + i\nuh \right)\left( \omega + i\nuhe \right) + \nuhhe\nuheh} \nonumber \\
&&+   \frac{\nuihe \nuhei \left( \omega + i\nuh \right) + i \nuihe \nuheh \nuhi}{\left( \omega + i\nuh \right)\left( \omega + i\nuhe \right) + \nuhhe\nuheh}. \label{eq:omegacol}
\end{eqnarray}
The effective density takes into account that the inertia of the plasma in response to the oscillations of the magnetic field depends on the coupling degree between the different ionized and neutral species. When the wave frequency is much higher than the collision frequencies between species, $\Omega_{\rm col} \to \omega$ so that $\rho_{\rm eff} \to \rhoi$, meaning that high-frequency Alfv\'en waves only perturb the ionic fluid while neutrals are left at rest.  Conversely, when the wave frequency is much lower than the collision frequencies, $\Omega_{\rm col} \to \omega\rho/\rhoi$ so that $\rho_{\rm eff} \to \rho$, meaning that low-frequency Alfv\'en waves  perturb the whole plasma and all species move as a single fluid. There is a continuous transition when the wave frequency varies between both limits.

Finally, we use Equation~(\ref{eq:vi}) in Equation~(\ref{eq:induction}) to get an equation involving $B'_\varphi$ alone, namely
\begin{eqnarray}
i\omega B'_\varphi &+& r {\bf B} \cdot \nabla \left[ \frac{i}{\omega} \frac{1}{\mu\rho_{\rm eff}} \frac{1}{r^2} {\bf B} \cdot \nabla \left( r B'_\varphi \right) \right] \nonumber \\
 &+& \eta \left(  \nabla^2 B'_\varphi - \frac{1}{r^2}  B'_\varphi\right) + \frac{\partial \eta}{\partial z} \frac{\partial B'_\varphi}{\partial z}  =0. \label{eq:main}
\end{eqnarray}
Equation~(\ref{eq:main}) is our main equation and gives us the spatial dependence of the torsional wave perturbations in the flux tube. We recall that this equation is obtained for the case that $\partial/\partial\varphi = 0$. Once the magnetic field perturbation is obtained by solving Equation~(\ref{eq:main}) along with appropriate boundary conditions, the ion velocity perturbation can be computed from Equation~(\ref{eq:vi}).

\subsection{Wave energy equation}
\label{sec:energy}

The linearized equations (Equations~(\ref{eq:momi})--(\ref{eq:induction})) can be manipulated  to obtain an equation describing the evolution of the energy of the waves  \citep[see][]{walker2005}, namely
\begin{equation}
\frac{\partial U}{\partial t} + \nabla \cdot {\bf \Pi} = - H, \label{eq:energycon}
\end{equation}
where $U$ is the total (kinetic + magnetic) energy density, ${\bf \Pi}$ is the energy flux, and $H$ is the loss of  energy owing to dissipation. These quantities are given by
\begin{eqnarray}
U  &=& \frac{1}{2} \rhoi \left| {\bf v'_{\rm i}} \right|^2 + \frac{1}{2} \rhoh  \left| {\bf v'_{\rm H}} \right|^2 + \frac{1}{2} \rhohe  \left| {\bf v'_{\rm He\,I}} \right|^2  \nonumber \\
&& + \frac{1}{2\mu}\left| {\bf B'} \right|^2 , \label{eq:energydensity} \\
{\bf \Pi}  &=& \frac{1}{\mu} \left[  \left({\bf B} \cdot {\bf B'} \right) {\bf v'_{\rm i}}  - \left({\bf v'_{\rm i}}  \cdot {\bf B'} \right){\bf B} \right] + \eta {\bf J'} \times {\bf B'}, \label{eq:energyflux} \\
H &=& H_{\rm Ohm.} + H_{\rm fric.} \label{eq:heat}
\end{eqnarray}
with
\begin{eqnarray}
H_{\rm Ohm.} &=& \mu\eta \left| {\bf J'}\right|^2 = \frac{\eta}{\mu} \left| \nabla \times {\bf B'} \right|^2, \\
H_{\rm fric.} &=& \aih \left| {\bf v'_{\rm i}} - {\bf v'_{\rm H}} \right|^2 +  \aihe \left| {\bf v'_{\rm i}} - {\bf v'_{\rm He\,I}}\right|^2 \nonumber \\
&& +  \ahhe \left| {\bf v'_{\rm H}} - {\bf v'_{\rm HeI}} \right|^2. 
\end{eqnarray}

We note that, because of total energy conservation, the energy dissipated from the waves must necessarily by converted into internal energy of the plasma, so that $H$ is the heating rate associated to wave dissipation. The two contributions to the heating rate are Ohmic heating, $H_{\rm Ohm.}$, owing to the dissipation of currents, and frictional heating, $H_{\rm fric.}$, caused by velocity drifts between species.  The role of $H_{\rm Ohm.}$ was not considered by \citet{soler2017} and may be important in the lower chromosphere where $\eta$ is large (see again Figure~\ref{fig:background}(c)).

The expression of the energy flux, ${\bf \Pi}$, contains the term $\eta {\bf J'} \times {\bf B'}$, which is absent from the formula used by \citet{soler2017} because they considered $\eta = 0$. However, this additional term can be neglected throughout the whole atmosphere. To show that, we introduce the quantities $B_0$ and $B_1$, with $B_1 \ll B_0$, representing the characteristic values of the background magnetic field strength and its perturbation, respectively, $v_1$ representing the characteristic amplitude of the ion velocity perturbation, and $\lambda_0$ representing the characteristic wavelength. We define $R_m = \lambda_0 v_1 /\eta$ as the effective magnetic Reynolds number associated to the waves. In the lower chromosphere where Ohm's diffusion is most efficient, $\eta \sim 10^4$~m$^2$~s$^{-1}$  (see Figure~\ref{fig:background}(c)), and we can assume $v_1 \sim 1$~km~s$^{-1}$ as an order-of-magnitude value of the velocity amplitude according to observations \citep[e.g.,][]{jess2009,Matsumoto2010}. These values result in $R_m \sim \lambda_0 / \left(10~\textrm{m}\right)$, which means that $R_m$ is a large parameter unless extremely short  wavelengths are considered. Then, by comparing the  magnitudes of the two terms in Equation~(\ref{eq:energyflux}), we find that their ratio is the product of two small quantities, $1/R_m$ and $B_1/B_0$, namely
\begin{equation}
\frac{\left| \eta {\bf J'} \times {\bf B'} \right|}{\left| \frac{1}{\mu} \left[  \left({\bf B} \cdot {\bf B'} \right) {\bf v'_{\rm i}}  - \left({\bf v'_{\rm i}}  \cdot {\bf B'} \right){\bf B} \right]\right|} \sim \frac{1}{R_m} \frac{B_1}{B_0} \ll 1.
\end{equation}
Therefore, we can safely neglect the term proportional to $\eta$ in the expression of the wave energy flux and use the same expression as in \citet{soler2017}, namely
\begin{equation}
{\bf \Pi}  \approx \frac{1}{\mu} \left[  \left({\bf B} \cdot {\bf B'} \right) {\bf v'_{\rm i}}  - \left({\bf v'_{\rm i}}  \cdot {\bf B'} \right){\bf B} \right] = -\frac{1}{\mu} v'_{\rm i,\varphi} B'_\varphi {\bf B}. \label{eq:energyflux2}
\end{equation}

\subsection{Energy propagation}
\label{sec:coefs}

Because of the  dependence $\exp \left( - i \omega t \right)$,  the energy flux,  ${\bf \Pi} $ is oscillatory in time. To avoid the oscillations and compute the net contribution, we average ${\bf \Pi}$  in time over one full period of the wave, $2\pi/\omega$. Then, the time-averaged  energy flux, $\left< {\bf \Pi} \right>$, is given by
\begin{equation}
\left< {\bf \Pi} \right> = -\frac{1}{2\mu} {\rm Re}\left( v'_{\rm i,\varphi} {B'_\varphi}^{*} \right) {\bf B}, \label{eq:flux} 
\end{equation}
where $*$ denotes the complex conjugate. The time-averaged energy flux, $\left< {\bf \Pi} \right>$, informs us about the net energy propagated by the Alfv\'en waves. However, as shown in \citet{soler2017}, the  perturbations in the flux tube are the result of the superposition of waves propagating parallel and anti-parallel to the magnetic field direction. In the case of the vertical flux tube, parallel/anti-parallel propagation essentially corresponds to upward/downward propagation. For the present study, it is crucial to distinguish between both directions of propagation and to separate the associated energy fluxes.

Inspired by the studies of Alfv\'enic turbulence, an adequate method that enables to split the Alfv\'en wave perturbations into the two possible directions of propagation involves the so-called Els\"asser variables \citep[see, e.g.,][]{biskamp2008}. Based on the classic Els\"asser variables, we define the modified Els\"asser variables for the multi-fluid case as
\begin{eqnarray}
Z^\uparrow &=& v'_{\rm i,\varphi} - \frac{1}{\sqrt{\mu \rho_{\rm eff}}} B'_\varphi, \\
Z^\downarrow &=& v'_{\rm i,\varphi} + \frac{1}{\sqrt{\mu \rho_{\rm eff}}} B'_\varphi,
\end{eqnarray}
where $Z^\uparrow$ describes Alfv\'en waves  propagating in the direction of $\bf B$ (upward propagation), $Z^\downarrow$ describes Alfv\'en waves  propagating opposite to $\bf B$ (downward propagation), and $\rho_{\rm eff}$ is the effective  density that was defined before (Equation~(\ref{eq:effectiverho})).

We express $v'_{\rm i,\varphi}$ and $B'_\varphi$ in terms of $Z^\uparrow$ and $Z^\downarrow$, and insert the resulting formulae into the expression of the time-averaged energy flux (Equation~(\ref{eq:flux})). Then, we find that $\left< {\bf \Pi} \right> = \left< {\bf \Pi} \right>^\uparrow + \left< {\bf \Pi} \right>^\downarrow$, with
\begin{eqnarray}
\left< {\bf \Pi} \right>^\uparrow &=& \frac{{\rm Re}\left(\sqrt{\rho_{\rm eff} }\right)}{8\sqrt{\mu}}Z^\uparrow Z^{\uparrow *}\, {\bf B}, \\
\left< {\bf \Pi} \right>^\downarrow &=& -\frac{{\rm Re}\left(\sqrt{\rho_{\rm eff} }\right)}{8\sqrt{\mu}}Z^\downarrow Z^{\downarrow *}\, {\bf B},
\end{eqnarray}
where $\left< {\bf \Pi} \right>^\uparrow$ and $\left< {\bf \Pi} \right>^\downarrow$ correspond to the time-averaged energy fluxes associated to the upward propagating and downward propagating Alfv\'en waves, respectively.

We are interested in the propagation of wave energy along the vertical direction. Therefore, to drop the other two coordinates, $r$ and $\varphi$, we horizontally average  the energy fluxes  at each height over an area extending from $r=0$ to $r=r_{\rm max}$, namely
\begin{equation}
\left< {\bf \Pi} \right>^{\uparrow\downarrow}_{\rm av.} = \frac{1}{\pi r_{\rm max}^2}\int_0^{2\pi} \int_0^{r_{\rm max}} \left< {\bf \Pi} \right>^{\uparrow\downarrow}  r {\rm d}r {\rm d}\varphi.
\end{equation}
The horizontally-averaged fluxes are functions of $z$ alone and can be used to quantify, at a certain height, the fractions of the total wave energy that reflect and propagate at that height.

On the other hand, we define the reflection, $\mathcal{R}$, and transmission, $\mathcal{T}$,  coefficients, which physically represent the fractions of the driven wave energy that are reflected back to the photosphere and are transmitted to the corona, respectively. These  coefficients of wave energy reflection and transmission are intrinsic properties of the background model as a whole, i.e., they are independent of $z$, but depend on the wave frequency \citep[see][]{soler2017}. Assuming that the waves are driven at the lower photospheric boundary, $z=z_{\rm ph}$, and that there are no incoming waves from the upper coronal boundary, $z=z_{\rm c}$, the incident, $\left< {\Pi} \right>_{\rm inc}$, reflected, $ \left< {\Pi} \right>_{\rm ref}$, and transmitted, $\left< {\Pi} \right>_{\rm tra}$,  fluxes at those boundaries are
\begin{eqnarray}
\left< {\Pi} \right>_{\rm inc} &=&  \left< {\bf \Pi} \right>^{\uparrow}_{\rm av.} \cdot \hat{e}_z, \qquad \textrm{at} \qquad z=z_{\rm ph}, \\
 \left< {\Pi} \right>_{\rm ref} &=&  \left< {\bf \Pi} \right>^{\downarrow}_{\rm av.} \cdot \hat{e}_z, \qquad \textrm{at} \qquad z=z_{\rm ph}, \\
 \left< {\Pi} \right>_{\rm tra} &=& \left< {\bf \Pi} \right>^{\uparrow}_{\rm av.} \cdot \hat{e}_z, \qquad \textrm{at} \qquad z=z_{\rm c}.
\end{eqnarray}
We note that only the $z$-components, i.e., the normal components, of $ \left< {\bf \Pi} \right>^{\uparrow\downarrow}_{\rm av.}$ at the boundaries are needed. The incident flux, $\left< {\Pi} \right>_{\rm inc}$, is imposed by the wave driver, while the reflected, $ \left< {\Pi} \right>_{\rm ref}$, and transmitted, $\left< {\Pi} \right>_{\rm tra}$,  fluxes depend on the reflective properties of the background atmospheric model and magnetic field and on the efficiency of the dissipation mechanisms. Then, we compute the  coefficients as 
\begin{equation}
\mathcal{R} = -\frac{\left< {\Pi} \right>_{\rm ref} }{\left< {\Pi} \right>_{\rm inc}}, \qquad \mathcal{T} = \frac{\left< {\Pi} \right>_{\rm tra} }{\left< {\Pi} \right>_{\rm inc}}.
\end{equation}
Furthermore, by  invoking conservation of energy, we can compute the fraction of the incident wave energy that is deposited  or absorbed in the  plasma because of dissipation, $\mathcal{A}$,  namely  
\begin{equation}
\mathcal{A} = 1 - \mathcal{R} - \mathcal{T}.
\end{equation}
Obviously, the absorption is also frequency-dependent because the efficiency of the dissipation mechanisms depends on $\omega$.

\subsection{Heating}
\label{sec:heats}

As in the case of the energy flux, the wave heating rate,  $H$, is oscillatory in time. As before, to calculate the net heating we compute the time-averaged heating rate, $\left< H \right>$, as
\begin{equation}
\left< H \right> = \left< H \right>_{\rm Ohm.} + \left< H \right>_{\rm fric.}, \label{eq:heatingav}
\end{equation}
where $\left< H \right>_{\rm Ohm.}$ and $\left< H \right>_{\rm fric.}$ represent the Ohmic and frictional contributions to the time-averaged heating and are given by
\begin{eqnarray}
\left< H \right>_{\rm Ohm.} &=& \frac{\eta}{2\mu} \left[ \frac{1}{r^2}\frac{\partial\left( r B'_\varphi \right)}{\partial r} \frac{\partial\left( r {B'_\varphi}^* \right)}{\partial r} \right. \nonumber \\
 && \left.+ \frac{\partial B'_\varphi}{\partial z} \frac{\partial {B'_\varphi}^*}{\partial z} \right], \\
\left< H \right>_{\rm fric.} &=& \frac{1}{2} \left( \aih \left| 1 - \frac{q_1}{p}  \right|^2 +  \aihe \left|  1 - \frac{q_2}{p} \right|^2 \right. \nonumber \\
&& \left. + \ahhe \left| \frac{q_1 - q_2}{p} \right|^2\right)  v'_{\rm i,\varphi} {v'_{\rm i,\varphi}}^*,
\end{eqnarray}
with
\begin{eqnarray}
p &=& \left( \omega + i\nuh \right)\left( \omega + i\nuhe \right) + \nuhhe\nuheh, \\
q_1 &=& i \nuhi \left( \omega + i\nuhe \right) - \nuhhe \nuhei, \\
q_2 &=& i \nuhei \left( \omega + i\nuh \right) - \nuheh \nuhi.
\end{eqnarray}

Finally, as for the case of the energy flux, we define the horizontally-averaged heating rate as 
\begin{equation}
\left< H \right>_{\rm av.} = \frac{1}{\pi r_{\rm max}^2}\int_0^{2\pi} \int_0^{r_{\rm max}} \left< H\right>  r {\rm d}r {\rm d}\varphi,
\end{equation}
which is a function of height alone.

\section{Numerical method of solution}
\label{sec:method}

The numerical scheme used to solve Equation~(\ref{eq:main})  is implemented in a {\em Wolfram Language} code run on {\em Mathematica} 11.3. We take advantage of the fact that the azimuthal direction, $\varphi$, is ignorable for torsional waves. Thus, Equation~(\ref{eq:main}) only needs to be integrated along the radial, $r$, and vertical, $z$, directions, while the solution is invariant in $\varphi$.  We consider a 2D numerical domain where $r\in\left[ 0, r_{\max} \right]$ and $z\in\left[ z_{\rm ph}, z_{\rm c}\right]$.  The code numerically solves Equation~(\ref{eq:main}) with the  function \verb&NDSolve&, using finite elements  for the spatial discretization   and considering appropriate boundary conditions at the ends of the numerical box (see Section~\ref{sec:bc}). In the code, the resolution of the numerical mesh is  nonuniform and is chosen to make sure that the solution is sufficiently accurate in physical terms (see Section~\ref{sec:resolution}).  After prescribing a value for the wave frequency, $f$, the output of the  integration routine is the spatial dependence of $B'_\varphi$ in the numerical domain.

 The result of waves excited by a broadband driver is constructed by varying $f$ in a wide range and superposing the perturbations obtained for individual frequencies within the range. The superposition is done after  assuming an spectral weighting function, a prescribed value of the incoming energy flux, and a random temporal phase at the photosphere.  The resulting total $B'_\varphi$ perturbation is used to compute all the other wave variables, including  the upward and downward energy fluxes and the heating rate.

\subsection{Boundary conditions}
\label{sec:bc}

Here we specify the boundary conditions used in the numerical code. The conditions at the lateral boundaries of the domain,  $r=0$ and $r=r_{\rm max}$, are set as
\begin{equation}
B'_\varphi = 0, \qquad \textrm{at} \qquad r=0 \quad \textrm{and} \quad r=r_{\rm max}.
\end{equation}
The condition at $r=0$ results from the mathematical requirement that Equation~(\ref{eq:main}) should remain finite at the center of the flux tube, while the condition at $r=r_{\rm max}$ results from the physical requirement that perturbations are confined within the flux tube.

Concerning the condition at the lower photospheric boundary, $z=z_{\rm ph}$, we assume that waves are driven just below that boundary and we can specify the form of the $B'_\varphi$. Hence, we take
\begin{equation}
B'_\varphi = A \left(f\right) b\left( r \right), \qquad \textrm{at} \qquad z=z_{\rm ph},
\end{equation}
where $A\left(f\right)$ is the spectral weighting function and $b\left( r \right)$ is an arbitrary function of $r$ that represents a twisting of the field lines at the photosphere. We assume that the perturbations at the photosphere are essentially confined within the flux patch, so that we take $b\left( r \right)$ of the form
\begin{equation}
b\left( r \right) =  r \exp\left[-\left( \frac{r}{R/2} \right)^2\right].
\end{equation}
We have tested other forms for the function $b\left( r \right)$, but no significant differences in the results are obtained. Naturally, the perturbations at the lower boundary are the superposition of incident (upward propagating) and reflected (downward propagating) waves. With the help of the modified Els\"asser variables, we can appropriately separate the upward and downward energy fluxes at the boundary regardless of the assumed form of $b\left( r \right) $. 

Regarding the spectral weighting function, we follow, e.g., \citet{tu2013} and \citet{arber2016} and take $A\left(f\right)$ in the form of a power-law dependence as
\begin{equation}
A\left(f\right) = A_0 \left\{ \begin{array}{lll}
\left(\frac{f}{f_p}\right)^{\epsilon_L}, & \textrm{if} & f \leq f_p, \\
\left(\frac{f}{f_p}\right)^{\epsilon_H}, & \textrm{if} & f > f_p,
\end{array} \right. \label{eq:spectral}
\end{equation}
where $f_p$ is the peak frequency of the spectrum, $\epsilon_L$ and $\epsilon_H$ are the low-frequency and high-frequency exponents, respectively, and $A_0$ is a constant that depends on the value of the injected energy flux averaged over the whole photosphere. In the computations we use $f_p = 1.59$~mHz and $\epsilon_H = -5/6$, while we consider two cases for $\epsilon_L$, namely $\epsilon_L = 0$ and $\epsilon_L = 5/6$. Thus, the spectrum consists of two regions: a low-frequency region where the power is either flat (for $\epsilon_L = 0$)  or  increases (for $\epsilon_L = 5/6$) with frequency, and a high-frequency region where the power decreases in a Kolmogorov-like fashion. As discussed by \citet{arber2016}, there is no direct observational evidence for such a driver spectrum for the waves, although the observed spectrum of photospheric horizontal velocities suggests a  power-law dependence \citep[see][]{Matsumoto2010}. The choice of a Kolmogorov spectrum at high frequencies is motivated by the observational indications that photospheric motions are turbulent. Regarding the form of the spectrum at low frequencies,  recent computations by \citet{vankooten2017} suggest a rather flat spectrum for low frequencies, i.e.,  $\epsilon_L \approx 0$, while \citet{tu2013} and \citet{arber2016} propose a low-frequency exponent of $\epsilon_L = 5/6$. These two possible choices for $\epsilon_L$ are considered to determine if this dependence has any impact on the results. 

We  consider that the photospheric driver contains a spectrum of frequencies ranging from $0.1$~mHz to 300~mHz, which corresponds to wave periods from 2.78~h to 3.33~s. The continuous spectrum is represented by 84 discrete frequencies with a logarithmic spacing. Recent observations of chromospheric torsional waves by \citet{Srivastava2017} reported frequencies in the range $12-42$~mHz, while \citet{jess2009} previously reported torsional waves in photospheric bright points with lower frequencies between $1-8$~mHz. Hence, the considered frequency range covers well the observed frequencies and extends the  range to lower and higher  values.

Regarding the injected energy flux at the photosphere, observations, analytic estimations, and numerical simulations indicate that the shaking \citep[see, e.g.,][]{Spruit1981,Choudhuri1993,Huang1995} and/or the twisting \citep[see, e.g.,][]{Shelyag2011,Shelyag2012,Wedemeyer2012,morton2013} of the footpoints of the magnetic field lines can efficiently generate incompressible transverse waves.  Calculations of the transverse wave energy flux generated by horizontal photopheric motions show that the flux generated within the flux tubes is $\sim 10^9$~erg~cm$^{-2}$~s$^{-1}$ \citep[see, e.g.,][]{Spruit1981,Ulmschneider2000,noble2003,Shelyag2011}. However, this driven flux needs to be averaged over the whole photosphere to determine the constant $A_0$ in the the spectral weighting function (Equation~(\ref{eq:spectral})), so that the magnetic field filling factor has to be taken into account. Thus, we write the horizontally averaged incoming flux at the photosphere as  \citep[see][]{Spruit1981}
\begin{equation}
\left< {\Pi} \right>_{\rm inc} =10^9 \times F, \qquad \textrm{at} \qquad z=z_{\rm ph}, \label{eq:injectedflux}
\end{equation}
given in erg~cm$^{-2}$~s$^{-1}$, where $F$ is the photospheric filling factor. In our model, the flux tube radius at the photosphere is $\sim 100$~km, while we assume that the flux tube merges with neighboring tubes at a radial distance of $\sim$~1,000~km. Hence the  filling factor is $F \sim 0.01$ \citep[see also][]{stenflo2000,solanki2000}. Therefore, from Equation~(\ref{eq:injectedflux}) we obtain an average injected flux of $10^7$~erg~cm$^{-2}$~s$^{-1}$, which coincides with the photospheric Alfv\'enic flux typically assumed in the recent literature \citep[e.g.,][]{depontieu2001,goodman2011,tu2013,arber2016}.

Finally, at the upper coronal boundary, $z=z_{\rm c}$, we impose the condition that there are no incoming waves from the corona and that the upward propagating waves get through the upper boundary without reflection. We note that the condition of no reflection is  strictly imposed at the upper boundary alone. The waves are allowed to naturally reflect in their way from the photosphere to the corona without any restriction. This upper boundary condition imposes that $\left< {\bf \Pi} \right>^\downarrow = \bf 0$ at $z=z_{\rm c}$. In terms of $B'_\varphi$, the  condition at the upper boundary translates into
\begin{equation}
\frac{1}{r}{\bf B} \cdot \nabla \left( r B'_\varphi  \right) = i \omega \sqrt{\mu \rho_{\rm eff}}\,  B'_\varphi, \qquad \textrm{at} \qquad z=z_{\rm c}. 
\end{equation}
This last expression can be simplified by taking into account that the plasma at the coronal boundary is fully ionized, i.e.,  $\rho_{\rm eff} = \rhoi$, and that the magnetic field is vertical and constant, i.e., ${\bf B} = B_{\rm c} \hat{e}_z$. Thus, the upper boundary condition simply becomes a condition for the normal derivative of $B'_\varphi$, namely
\begin{equation}
\frac{\partial B'_\varphi}{\partial z} = i \frac{\omega}{v_{\rm A,c}}  B'_\varphi, \qquad \textrm{at} \qquad z=z_{\rm c},
\end{equation}
where $v_{\rm A,c} = B_c /\sqrt{\mu\rhoi}$ is the coronal Alfv\'en velocity.

We note that in \citet{soler2017} the treatment of the boundary conditions involved the use of {\em ghost cells} at the top and bottom boundaries where the inward and outward waves were analytically expressed in the form of plane waves. The present treatment based on the decomposition between upward and downward fluxes via the modified Els\"asser variables is more general and does not require the use of {\em ghost cells}. Nevertheless, the upper boundary condition obtained with the present method turns out to be exactly the same as that used in \citet{soler2017}. When the magnetic field is uniform at the boundary, both methods provide the same conditions. However, the present approach provides a more accurate decomposition between incident and reflected waves at the lower photospheric boundary, where the  magnetic field is not uniform.

\subsection{Numerical resolution}
\label{sec:resolution}

Here we address the issue of what numerical resolution is needed to obtain physically meaningful solutions to Equation~(\ref{eq:main}). An insufficient numerical resolution would result in a poor description of the spatial scales associated to the waves, which would directly affect the computations of the wave energy flux and heating rate.  Therefore, we need to 
adjust the spatial resolution to minimize numerical errors.

As a zeroth-order approximation, we can perform a local analysis of Equation~(\ref{eq:main}) by ignoring the spatial variations of the background quantities. By doing so, we can locally derive a  characteristic spatial scale that plays the role of the effective wavelength of the perturbations along the magnetic field direction, namely 
\begin{equation}
\lambda_{\rm eff}  \sim  \frac{v_{\rm A,eff}}{f},
\end{equation}
where $v_{\rm A,eff} = \left|{\bf B}\right|/\sqrt{\mu\rho_{\rm eff}}$ is the effective Alfv\'en velocity computed with the effective density.  To make sure that the numerical solution of Equation~(\ref{eq:main}) is sufficiently accurate, the resolution of the numerical mesh should be a fraction of $\lambda_{\rm eff}$. The numerical resolution  becomes a practical problem for high wave frequencies for which $\lambda_{\rm eff}$ is very short in the lower chromosphere. For the highest frequencies considered, the mesh needs to be so fine as to resolve wavelengths as small as 100~m, approximately. However, $\lambda_{\rm eff}$ increases rapidly with height because of the decrease of the density. Hence, the required numerical resolution varies with height by several orders of magnitude. Using a constant resolution results in   long execution times of the finite-element solver, so that a more convenient spatially-dependent resolution is used instead.

We have implemented a nonuniform mesh with a height-dependent local resolution equal to $\lambda_{\rm eff}/3$, approximately. Convergence test have shown that this resolution is enough for the solution to be sufficiently accurate and keeps the execution time within reasonable bounds. The local resolution increases with height, but we have imposed that the resolution saturates to 10~km at the height where that particular value is attained.  Then, from that height up to the coronal boundary, the mesh is uniform with a constant resolution of 10~km. We note that 10~km is much smaller than the actual resolution required to resolve the wavelengths in the coronal part of the domain. The reason for using a finer   mesh is that it provides more accurate values of the spatial derivatives of  $B'_\varphi$ in the coronal part of the domain, which is essential for the computation of the velocity perturbation and determination of the transmission coefficient at the upper boundary.

\section{Analysis of Results}
\label{sec:results}

Here we show and discuss the results of the numerical computations for the case of  a photospheric field strength of $B_{\rm ph} = 1$~kG and  a low-frequency exponent of $\epsilon_L = 5/6$. Unless otherwise stated, these parameters are used hereafter.

\subsection{Magnetic field and velocity perturbations}

\begin{figure*}
\epsscale{0.7}
\plotone{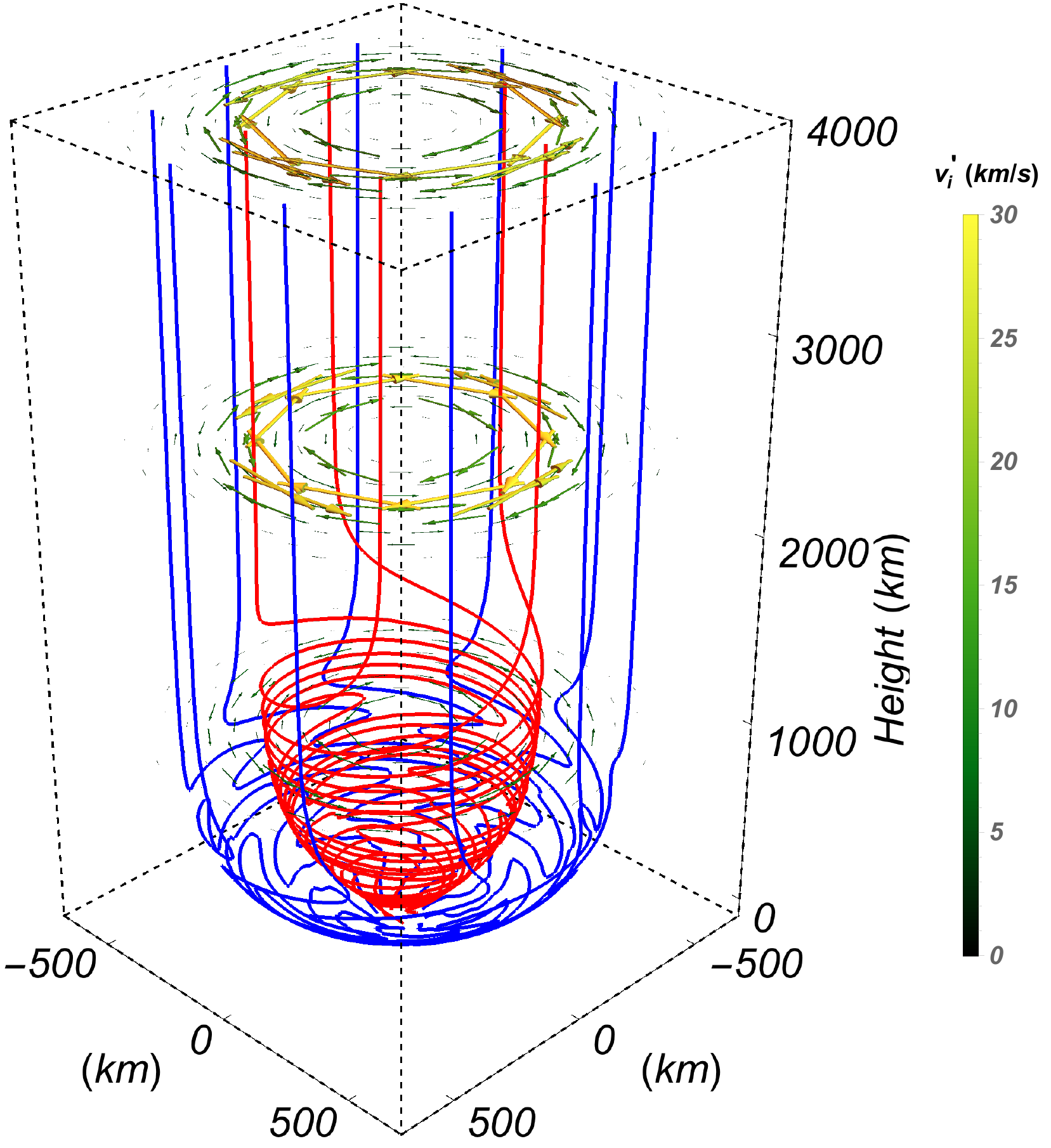}
\caption{Visualization in 3D of the  magnetic field and velocity perturbations associated to the torsional Alfv\'en waves excited by a photospheric broadband driver. The red and blue lines outline some selected magnetic field lines that cross the photosphere at $r=0.4 R = 40$~km and $r=0.7 R = 70$~km, respectively. The  arrows correspond to the ion velocity field at three horizontal planes located at 1,000~km, 2,500~km, and 4,000~km above the photosphere. The color and length of the arrows translate into the following speed values: short dark green  arrows for small velocites ($\lesssim 5$~km~s$^{-1}$),  intermediate bright green  arrows for moderate velocites ($\sim 10-15$~km~s$^{-1}$), and  long yellow arrows for large velocites ($\gtrsim 20$~km~s$^{-1}$). Results with $B_{\rm ph} = 1$~kG and $\epsilon_L = 5/6$.  
\label{fig:example}}
\end{figure*}

To start with, we display the magnetic field and velocity perturbations excited by the broadband driver. Figure~\ref{fig:example} shows a 3D view of the perturbations in the flux tube. This figure can be compared with Figure~\ref{fig:model}(a) corresponding to the undisturbed flux tube. We have selected some magnetic field lines (the same as those plotted in Figure~\ref{fig:model}(a)) and have computed their deformations because of the passing of the torsional Alfv\'en waves.  The field lines twist because of the torsional wave amplitude  is largest near the photosphere and decreases with height. The field lines remain practically undisturbed, i.e., straight, in the coronal part of the model, suggesting that most of wave power is not able to reach those heights. Later, we shall confirm this initial thought.

 In Figure~\ref{fig:example} we also represent the ion velocity perturbation, $v'_{\rm i,\varphi}$, at three horizontal planes located at 1,000~km, 2,500~km, and 4,000~km above the photosphere. It is clear that at those horizontal planes the plasma motions are torsional, i.e., polarized in the azimuthal direction. Contrary to the magnetic field perturbation, the amplitude of the velocity perturbation increases with height and the largest velocities are found in the coronal part.

\begin{figure*}
\epsscale{0.99}
\plottwo{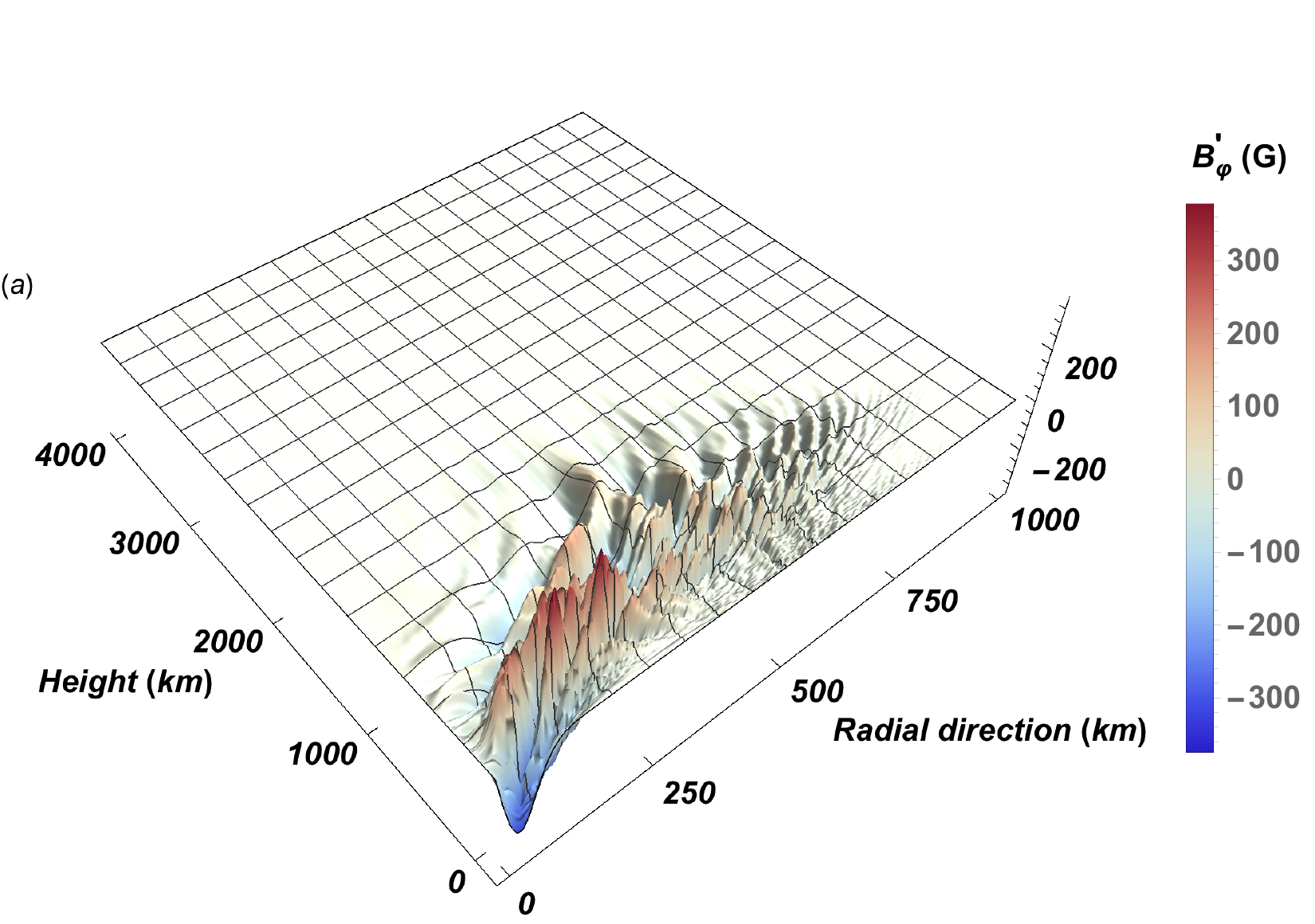}{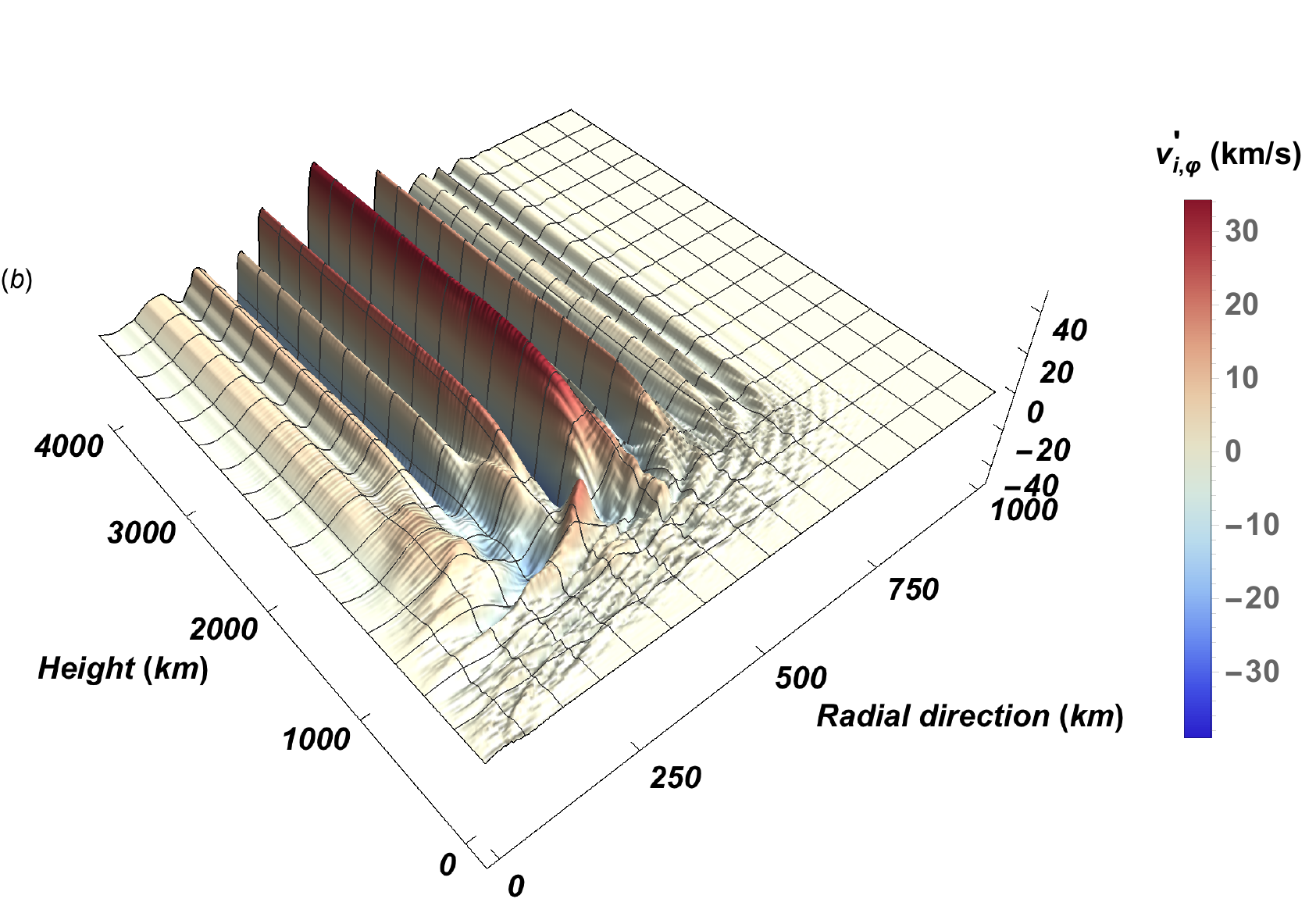}
\plottwo{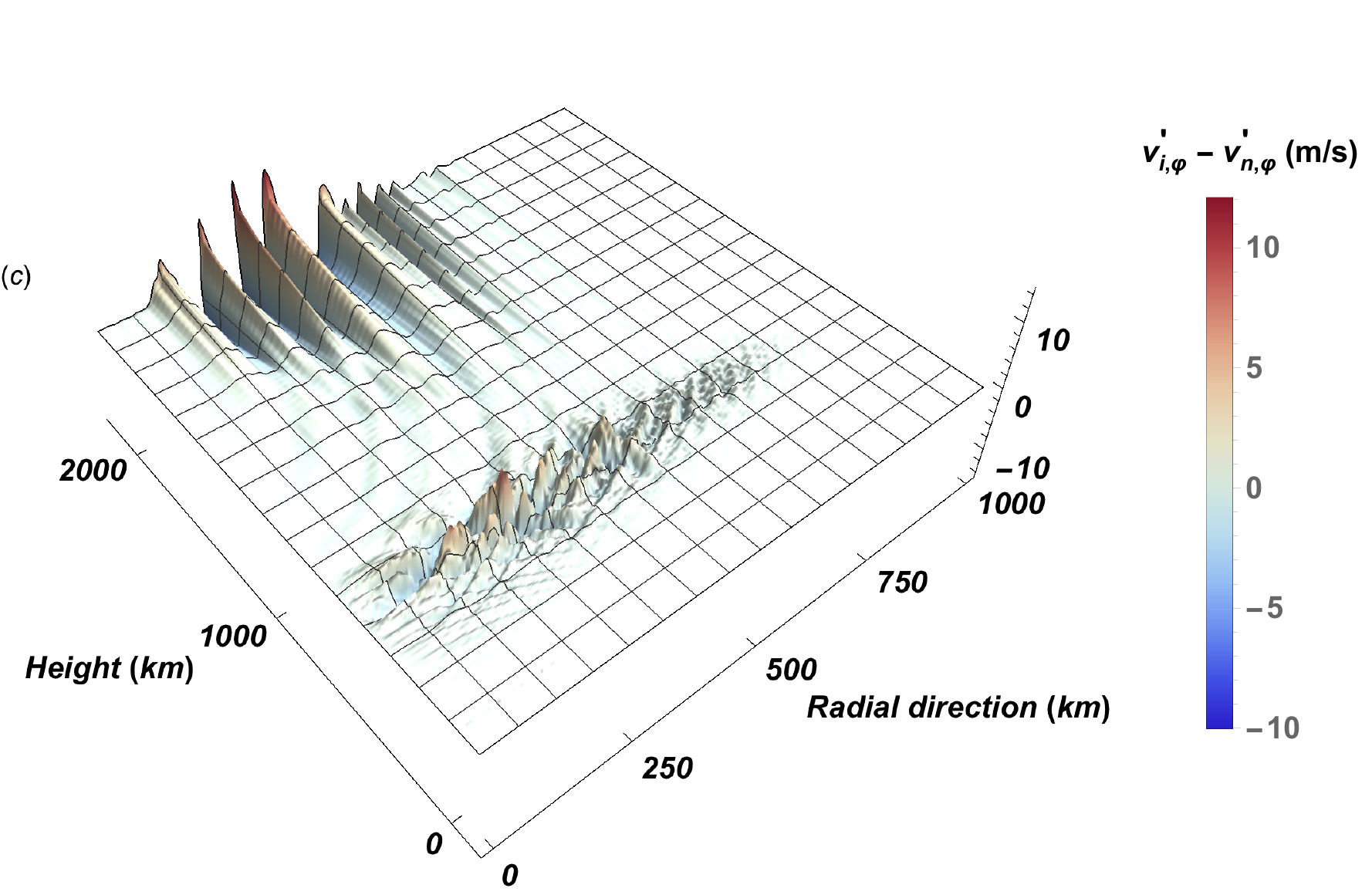}{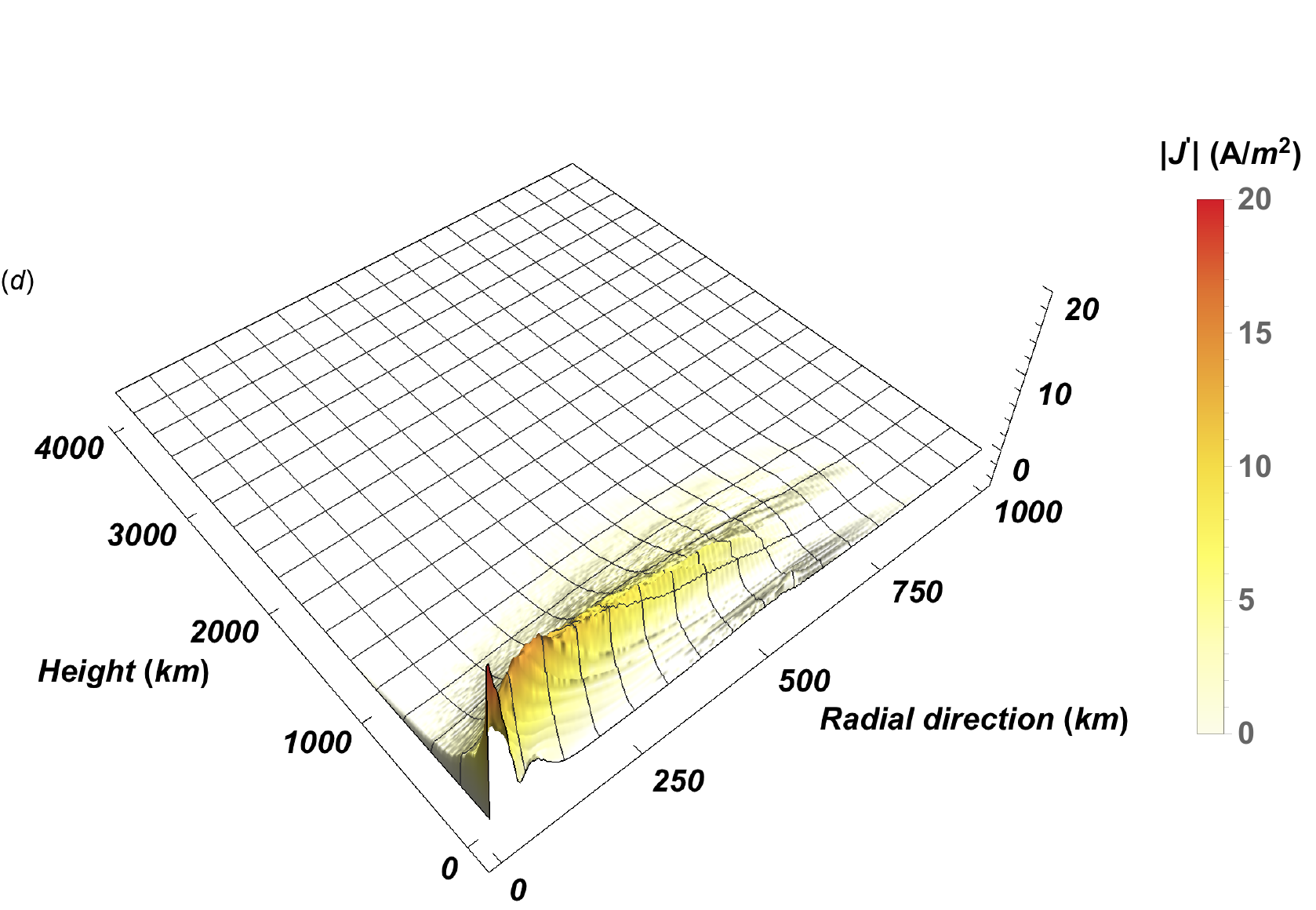}
\caption{Surface plots in the $rz$-plane of the (a) magnetic field perturbation, $B'_\varphi$,  (b) ion velocity perturbation, $v'_{\rm i,\varphi}$,  (c)  ion-neutral drift, $v'_{\rm i,\varphi} - v'_{\rm n,\varphi}$, and (d) modulus of the current density perturbation, $\left|{\bf J'}\right|$. Results with $B_{\rm ph} = 1$~kG and $\epsilon_L = 5/6$. Only the real part of the perturbations is plotted. Note that the ion-neutral drift is only plotted up to the transition region where the plasma gets fully ionized.
\label{fig:perts}}
\end{figure*}

\begin{figure*}
\plottwo{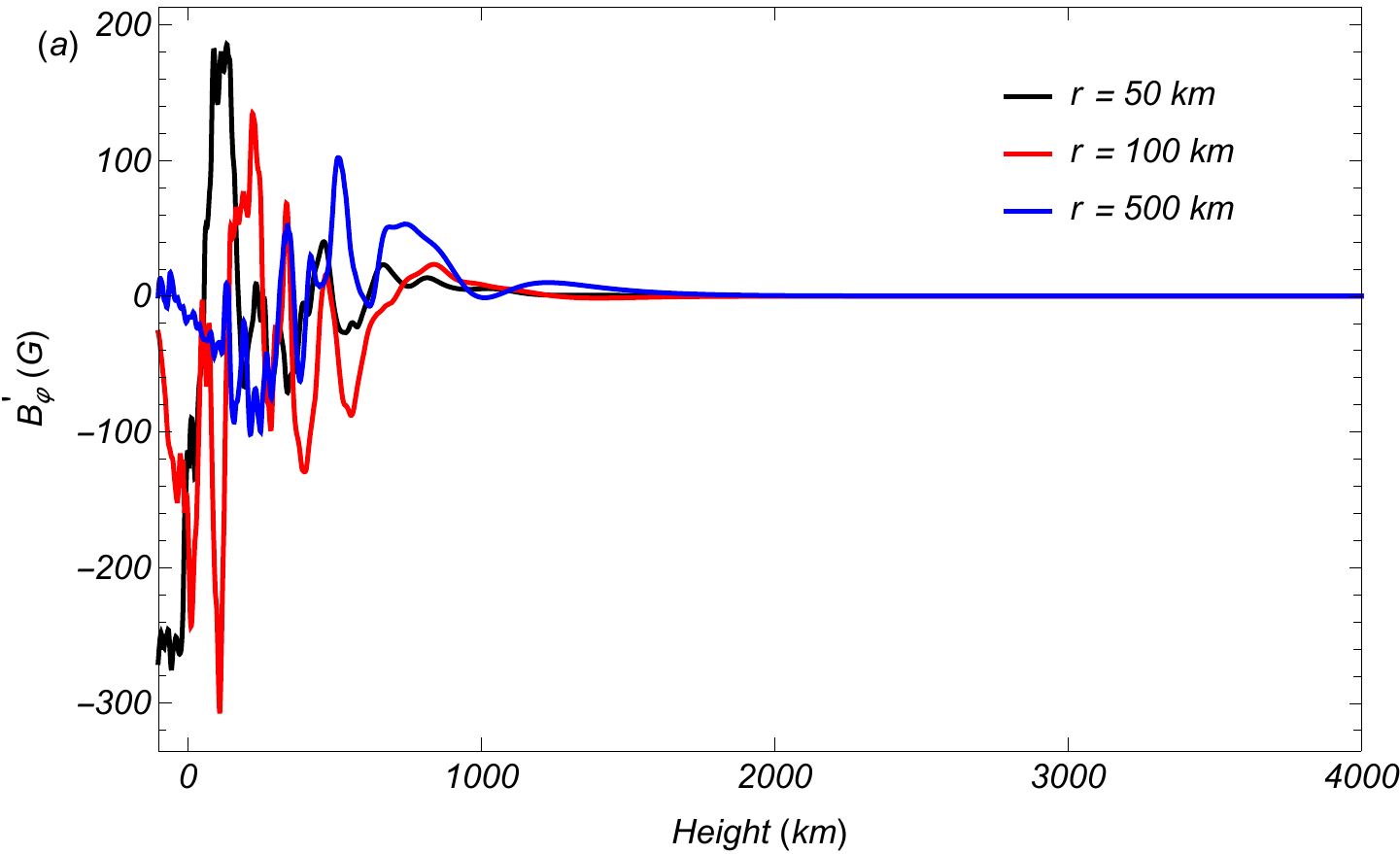}{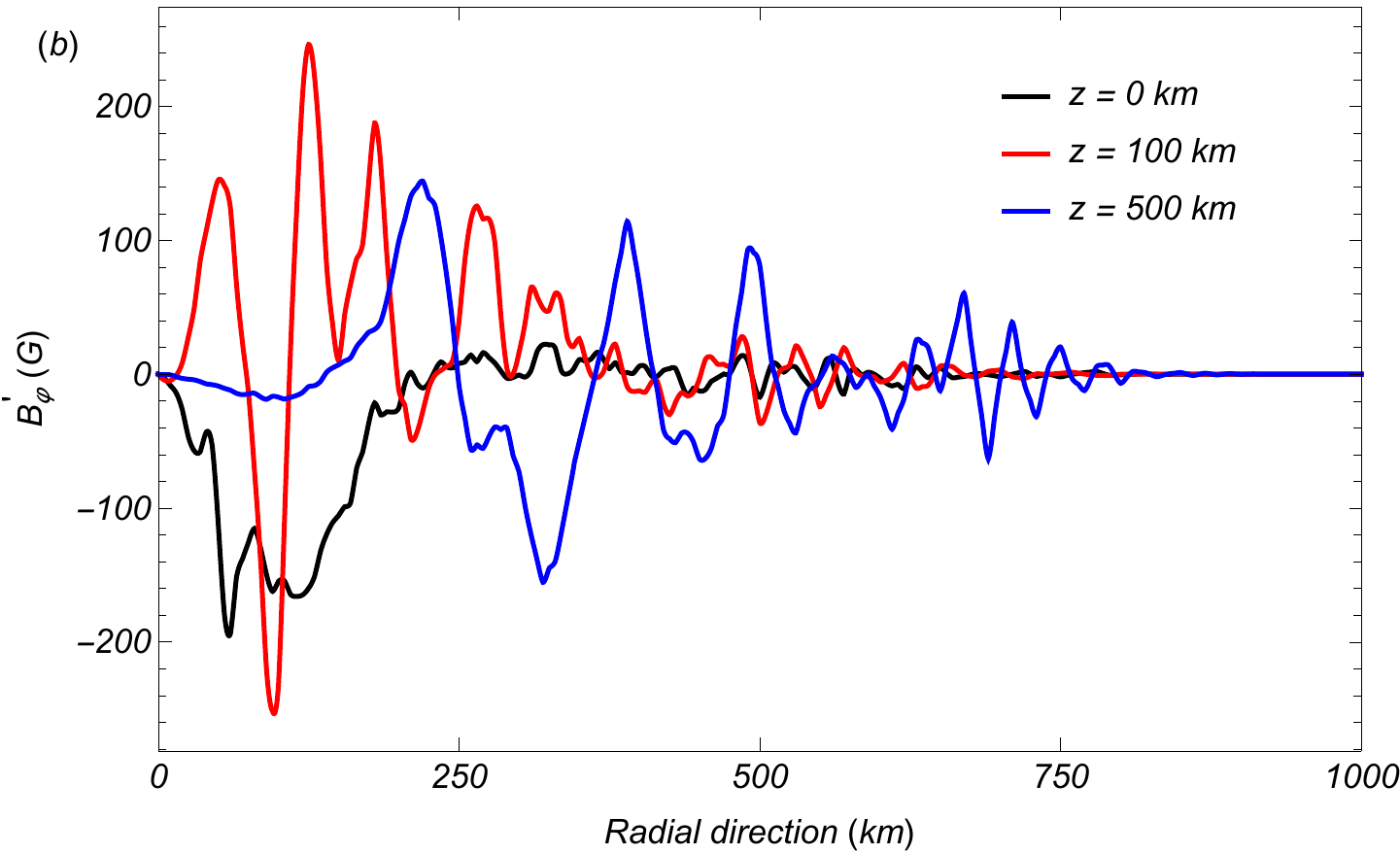}
\caption{Magnetic field perturbation, $B'_\varphi$. (a) Vertical cuts   at $r=50$~km, $r=100$~km, and $r=500$~km from the tube axis. (b) Horizontal cuts  at $z=0$~km, $z=100$~km,  and $z=500$~km above the photosphere.  Results with $B_{\rm ph} = 1$~kG and $\epsilon_L = 5/6$. Only the real part of the perturbation is plotted.
\label{fig:cutsbf}}
\end{figure*}

\begin{figure*}
\plottwo{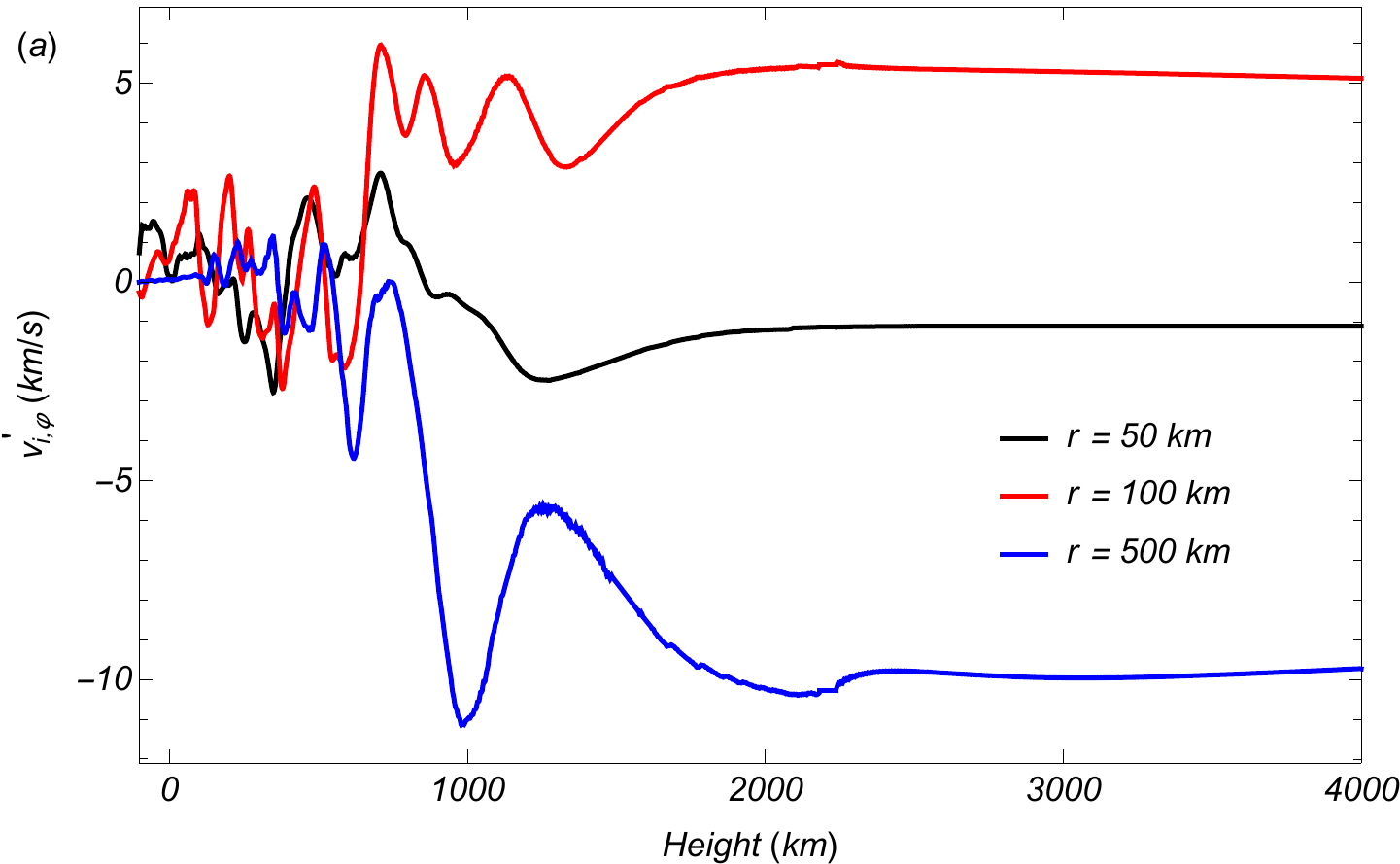}{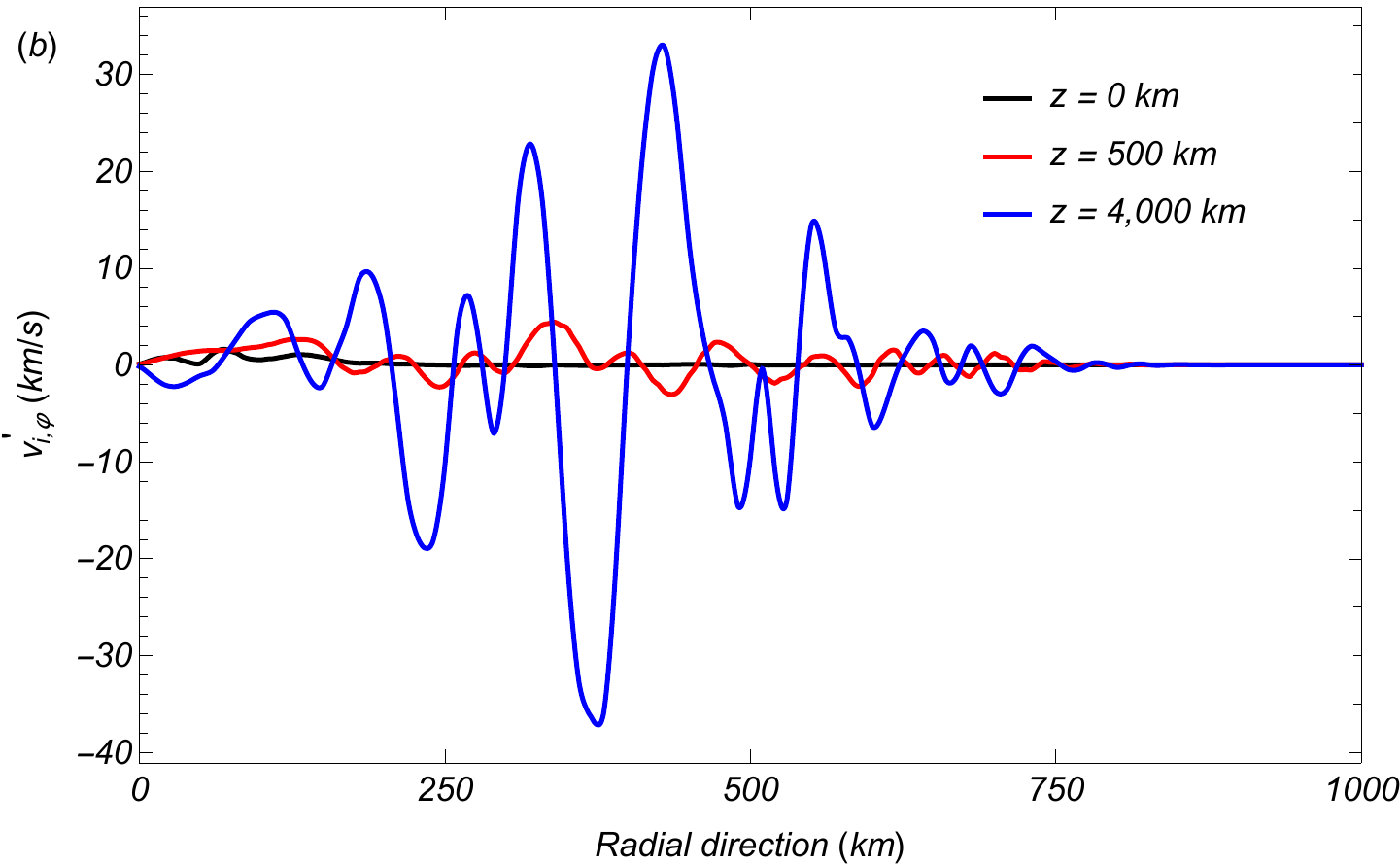}
\caption{Ion velocity perturbation, $v'_{\rm i,\varphi}$. (a) Vertical cuts at $r=50$~km, $r=100$~km, and $r=500$~km from the tube axis. (b) Horizontal cuts  at $z=0$~km, $z=500$~km,  and $z=4,000$~km above the photosphere.   Results with $B_{\rm ph} = 1$~kG and $\epsilon_L = 5/6$. Only the real part of the perturbation is plotted.
\label{fig:cutsvf}}
\end{figure*}

Since the torsional  perturbations are invariant in the $\varphi$-direction, a more illustrative way to show their shape is to remove the azimuthal dependence and plot the perturbations in the $rz$-plane. Figure~\ref{fig:perts} displays surface plots of the magnetic field perturbation, $B'_\varphi$, the ion velocity perturbation, $v'_{\rm i,\varphi}$, the ion-neutral drift, $v'_{\rm i,\varphi} - v'_{\rm n,\varphi}$, and the modulus of the current density perturbation, $\left|{\bf J'}\right|$. We have defined $v'_{\rm n,\varphi}$ as the neutrals center-of-mass velocity perturbation, which is computed as
\begin{equation}
v'_{\rm n,\varphi} = \frac{\rho_{\rm H} v'_{\rm H,\varphi} +\rho_{\rm He\,I} v'_{\rm He\,I,\varphi} }{\rho_{\rm H} + \rho_{\rm He\,I}}.
\end{equation}
In turn, Figures~\ref{fig:cutsbf}, \ref{fig:cutsvf}, \ref{fig:cutsvfn}, and \ref{fig:cutscurrent} show some selected vertical and horizontal cuts of those perturbations. The following discussion is based on the results displayed in these figures.

The amplitude of the magnetic field perturbation at the photospheric level is $\sim 200$~G, which is a relatively small fraction of the background photospheric field strength in this case, namely 1~kG. As  Figure~\ref{fig:example} suggested, now we clearly see that the magnetic field perturbation is essentially confined to low heights in the chromosphere and its amplitude decreases rapidly with height (see Figure~\ref{fig:cutsbf}(a)), while the opposite behavior is found in the case of the ion velocity perturbation (see Figure~\ref{fig:cutsvf}(a)). Essentially, this is the same result as that found in the 1.5D case of \citet[see their Figure~3]{soler2017}, although in the present 2.5D case the decrease of $B'_\varphi$ with height seems to be faster, suggesting a stronger damping. This fact will be  confirmed later. 

 The amplitude of the ion velocity perturbation is  $\sim 1-2$~km~s$^{-1}$ at the photosphere. The photospheric torsional velocities obtained here are similar to the velocity amplitudes of the torsional Alfv\'en waves in a bright point observed by \citet{jess2009} with SST/SOUP ($\sim 2.6$~km~s$^{-1}$)  and are also compatible with the quite-Sun photospheric horizontal velocities observed by \citet{Matsumoto2010} with  Hinode/SOT  ($\sim 1$~km~s$^{-1}$). The obtained velocity amplitudes in the chromosphere ($\lesssim 10$~km~s$^{-1}$) are of the same order of the chromospheric torsional motions reported by \citet{depontieu2014} with IRIS ($10-30$~km~s$^{-1}$). In our computations, the velocity amplitude increases to $\sim 20-40$~km~s$^{-1}$ when transition region and coronal heights are reached. These amplitudes agree well with the amplitudes of the outward-propagating Alfv\'enic waves observed at the transition region by \citet{mcintosh2011} with SDO/AIA ($\sim 20$~km~s$^{-1}$), although we note that the observations of \citet{mcintosh2011} could be better interpreted as Alfv\'enic kink waves.

We note, however, that these amplitudes depend on the value of the injected energy flux by the photospheric driver, so that increasing/decreasing the incoming flux would result in larger/smaller amplitudes.

\subsection{Ion-neutral drift}

\begin{figure*}
\plottwo{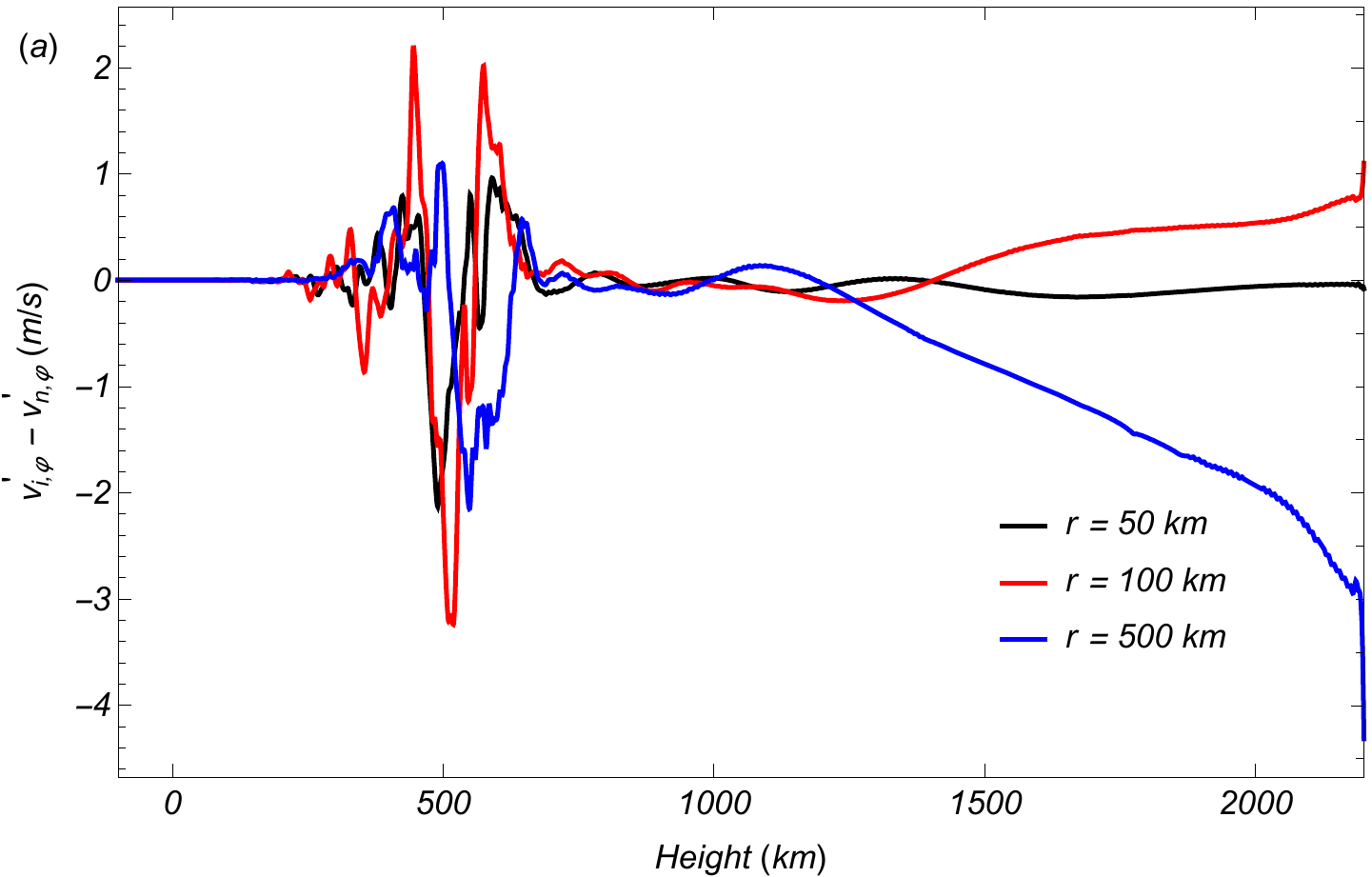}{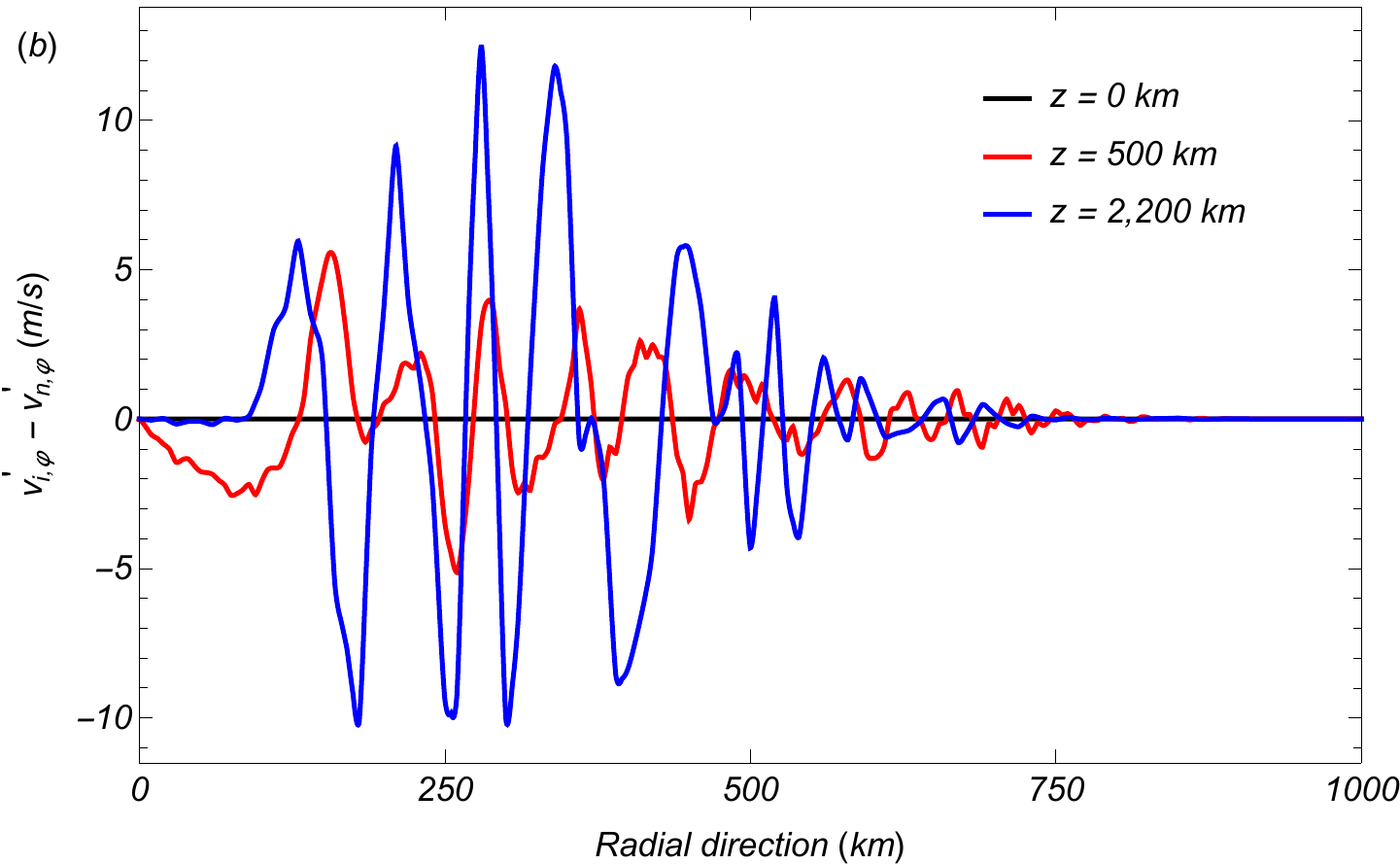}
\caption{Ion-neutral drift, $v'_{\rm i,\varphi} - v'_{\rm n,\varphi}$. (a) Vertical cuts  at $r=50$~km, $r=100$~km, and $r=500$~km from the tube axis. The horizontal axis (height) stops at the transition region, where the plasma becomes fully ionized. (b) Horizontal cuts at $z=0$~km, $z=500$~km,  and $z=2,200$~km above the photosphere.   Results with $B_{\rm ph} = 1$~kG and $\epsilon_L = 5/6$. Only the real part of the perturbation is plotted.
\label{fig:cutsvfn}}
\end{figure*}

Interestingly, the ion-neutral drift, $v'_{\rm i,\varphi} - v'_{\rm n,\varphi}$, displays a rather remarkable behavior with height (see Figure~\ref{fig:cutsvfn}). We remind readers that the ion-neutral drift is a measure of the strength of the coupling between ions and neutrals. Close to the photosphere, ions and neutrals are strongly coupled due to the very large density and negligible drifts are obtained. Although at those low heights the plasma is very weakly ionized, neutrals are so  tight to ions that all species move essentially as a single fluid following the magnetic field perturbations. However, the ion-neutral coupling becomes weaker as height increases. There is a narrow layer, centered around 500~km above the photosphere, where the ion-neutral drift suddenly increases to values of the order of $\sim 2-3$~m~s$^{-1}$. In that chromospheric layer, the ion-neutral drift is sufficiently large to produce an appreciable increase of the frictional heating (see Section~\ref{sec:heating}). Then, as height keeps increasing the ion-neutral drift  decreases again until the transition region is reached at about 2,200~km above the photosphere. The plasma gets fully ionized at the transition region, so that the  abundance of neutrals decreases dramatically at that height. Precisely,  it is just below the transition region that the ion-neutral drift takes its largest amplitudes of $\sim 10$~m~s$^{-1}$.  We shall see in Section~\ref{sec:heating} that the height-dependent behavior of the ion-neutral drift consistently explains the efficiency of the frictional heating in the chromosphere. The larger the ion-neutral drifts, the larger the heating rates. 

\subsection{Phase mixing and enhanced magnetic diffusion}

\begin{figure*}
\plottwo{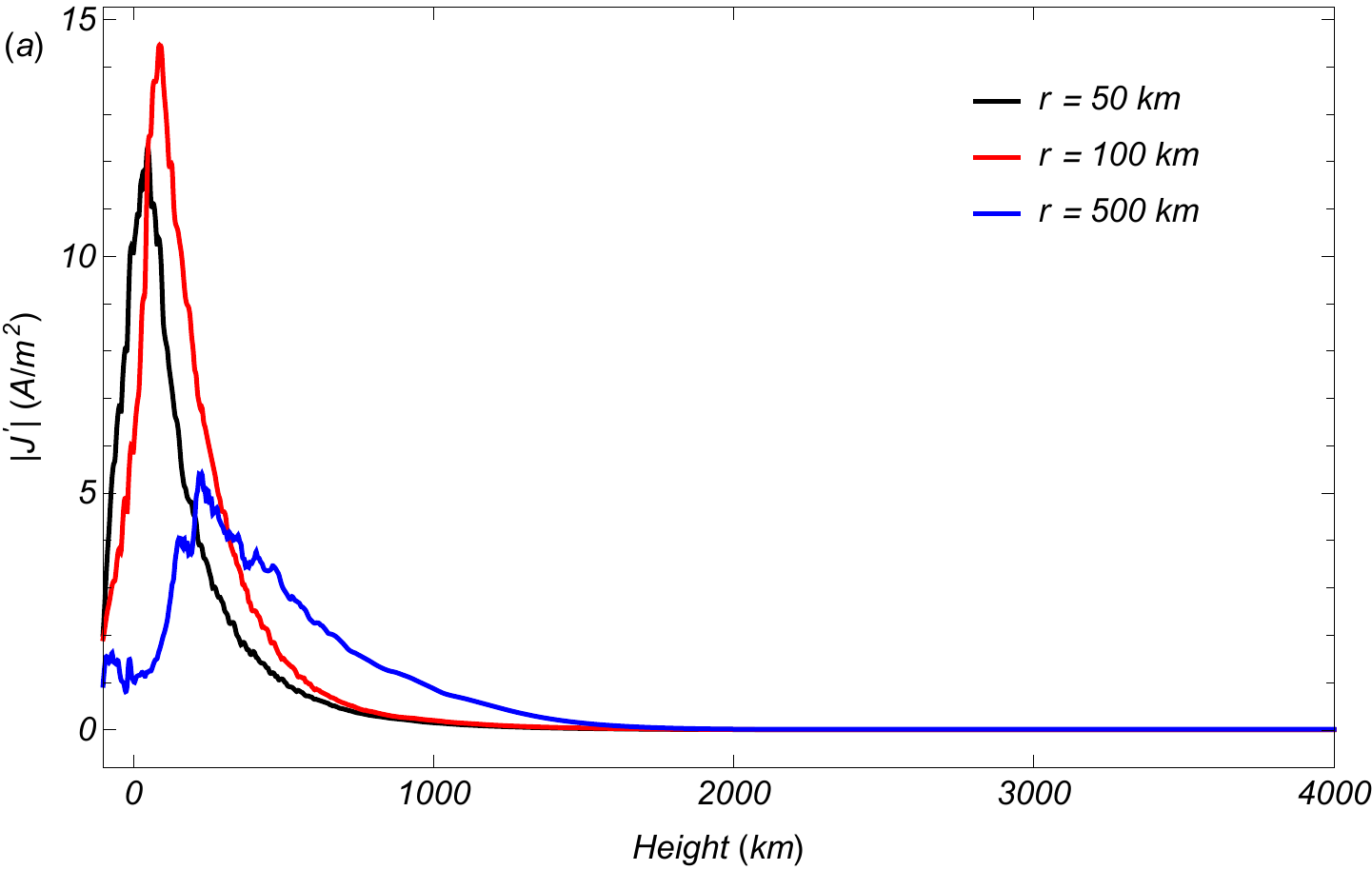}{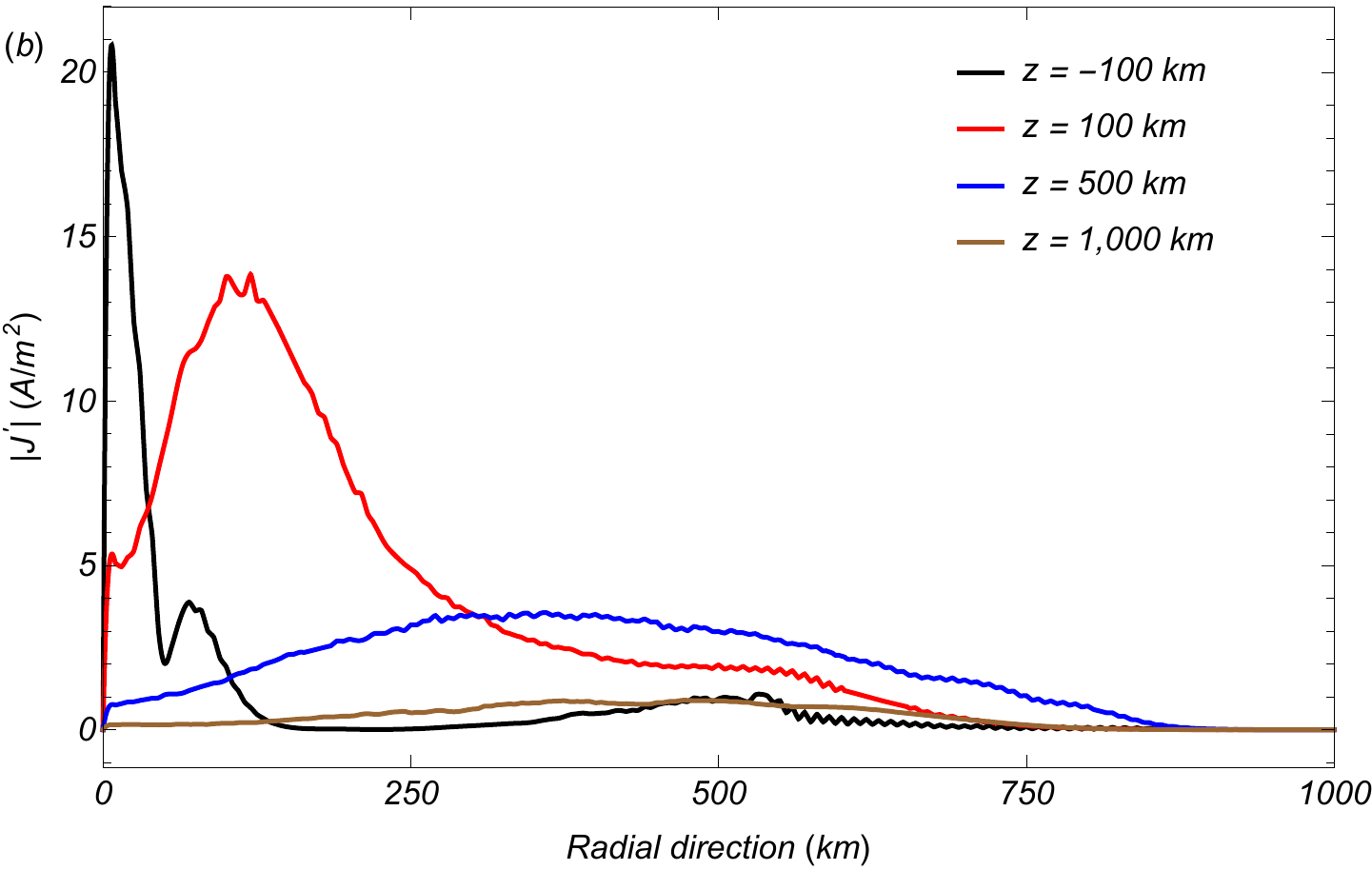}
\caption{Modulus of the current density perturbation, $\left|{\bf J'}\right|$. (a) Vertical cuts at $r=50$~km, $r=100$~km, and $r=500$~km from the tube axis. (b) Horizontal cuts  at $z=-100$~km, $z=100$~km, $z=500$~km, and $z=1,000$~km above the photosphere.  Results with $B_{\rm ph} = 1$~kG and $\epsilon_L = 5/6$.
\label{fig:cutscurrent}}
\end{figure*}

The plots corresponding to the current density perturbation (Figure~\ref{fig:cutscurrent}) show that the current is mainly localized in the photosphere and lower chromosphere and decreases rapidly with height. Practically, the  current density is completely damped for heights larger than 1,500~km, approximately. The reason for this strong damping of the current is Ohm's magnetic diffusion. Figure~\ref{fig:cutscurrent}(b), which displays some horizontals cuts of the current density at various heights, provides evidence that magnetic diffusion is at work. As height increases, the amplitude of the current decreases and, concurrently, the current spreads over a larger area across the flux tube.  

The horizontal cuts of the magnetic field and ion velocity perturbations (see Figures~\ref{fig:cutsbf}(b) and \ref{fig:cutsvf}(b)) reveal another important feature that directly affects the efficiency of Ohm's diffusion. These plots show that the perturbations develop small scales across the magnetic flux tube, specially at low heights in the chromosphere. This is especially evident in the case of the ion velocity perturbations. We interpret this shear in the perturbations as a clear evidence of the mechanism of phase mixing.

Phase mixing is an inherent process of Alfv\'en waves propagating in a structure with a gradient of the Alfv\'en velocity across the magnetic field direction \citep[see, e.g.,][]{Heyvaerts1983,nocera1984}. In our model, although the background density is  constant across the tube,  the magnetic field  expands horizontally so that its strength depends on the radial coordinate. This results in a radially varying Alfv\'en velocity. Because of the spatially dependent Alfv\'en velocity, waves propagating on adjacent field lines get out of phase as height increases, producing a  shear in the velocity and magnetic field perturbations across the magnetic field. In turn, this magnetic shear locally generates currents that enhance the efficiency of Ohm's diffusion, giving rise to a stronger damping of the waves.

It has been shown, both analytically and numerically \citep[see, e.g.,][]{ruderman1998,demoortel2000,smith2007,ruderman2017,petrukhin2018}, that divergence of the magnetic field lines enhances the efficiency of phase mixing, whereas gravitational stratification diminishes its effect compared to the case with no stratification. In our model, magnetic field lines heavily expand (i.e., they diverge) at low heights in the chromosphere. It is precisely at those low heights that phase mixing and magnetic diffusion are greatly enhanced, giving rise to a strong wave damping. Conversely, as height increases the field lines become nearly vertical, while density keeps decreasing due to stratification. Then, at large heights phase mixing becomes less efficient because the dominant effect is density stratification. In fact, it can be seen in the 2D plot of the ion velocity perturbation (Figure~\ref{fig:perts}(b)) that the generation of shear and small scales in the radial direction takes place at low heights predominantly. Then, as height increases, the radial structure of the perturbations remains practically unmodified in the upper part of the  domain. Therefore, our results agree well with the  behavior of Alfv\'en waves propagating in a magnetically divergent and gravitationally stratified medium explored in previous works in the literature \citep[see, e.g.,][]{smith2007}.

The role of Ohm's diffusion, enhanced via phase mixing, explains why in the present results the damping of the perturbations appears to be stronger than in \citet{soler2017}, where Ohm's diffusion was ignored. Phase mixing could not happen either in the case of  \citet{soler2017} because their model was 1.5D and the dependence across the magnetic field was not explicitly solved.

\subsection{Energy fluxes}

\begin{figure*}
\plottwo{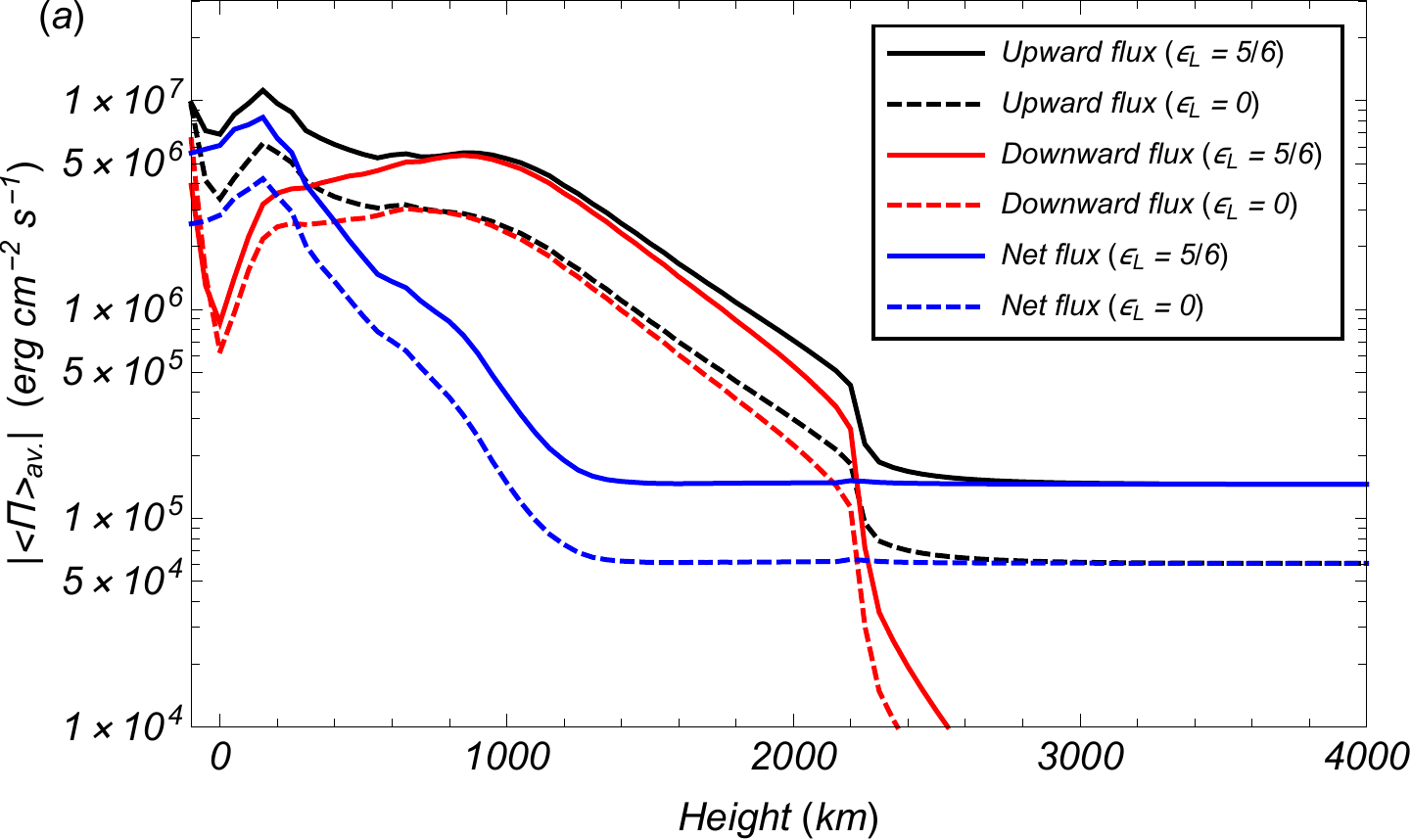}{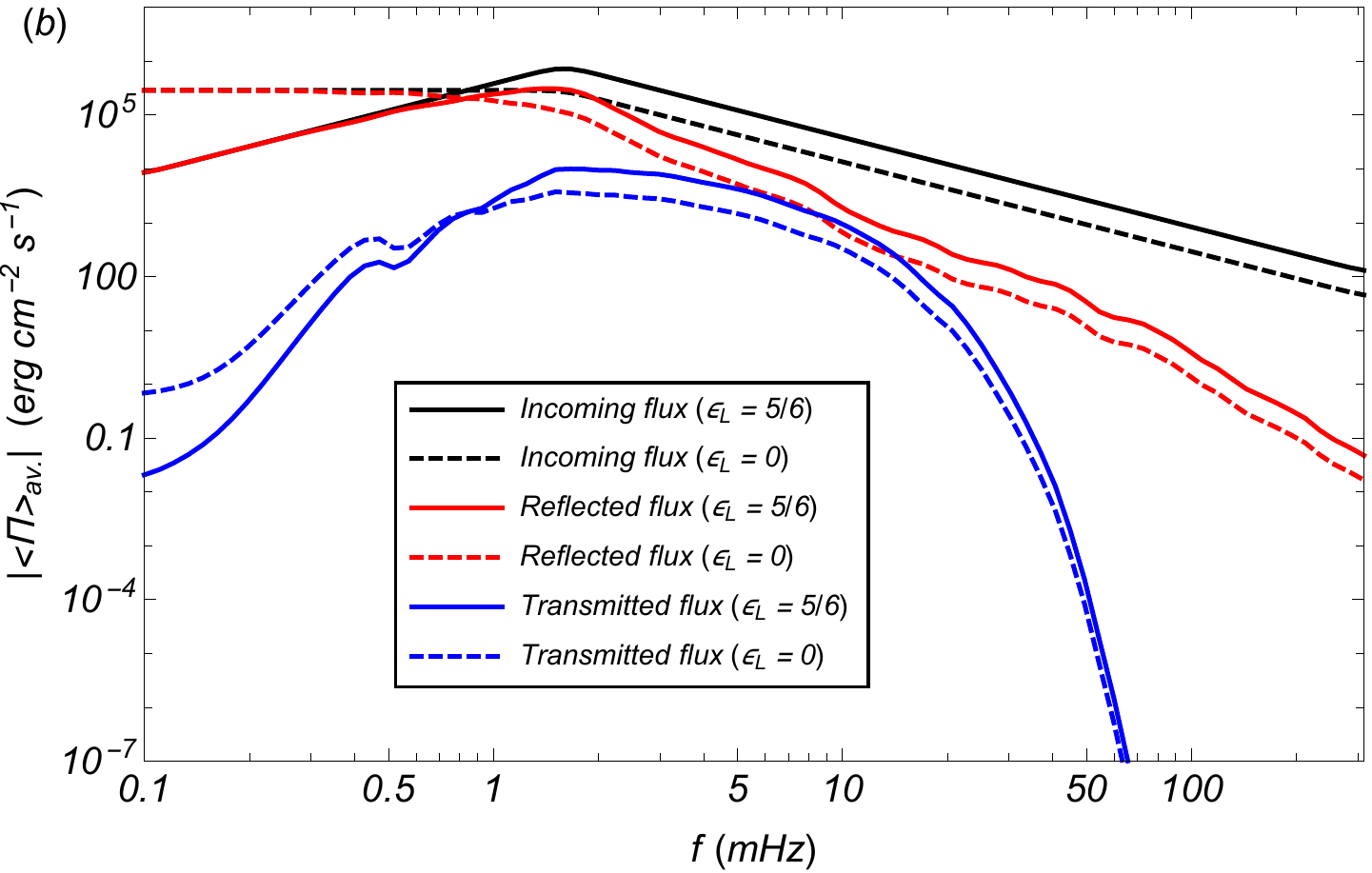}
\plottwo{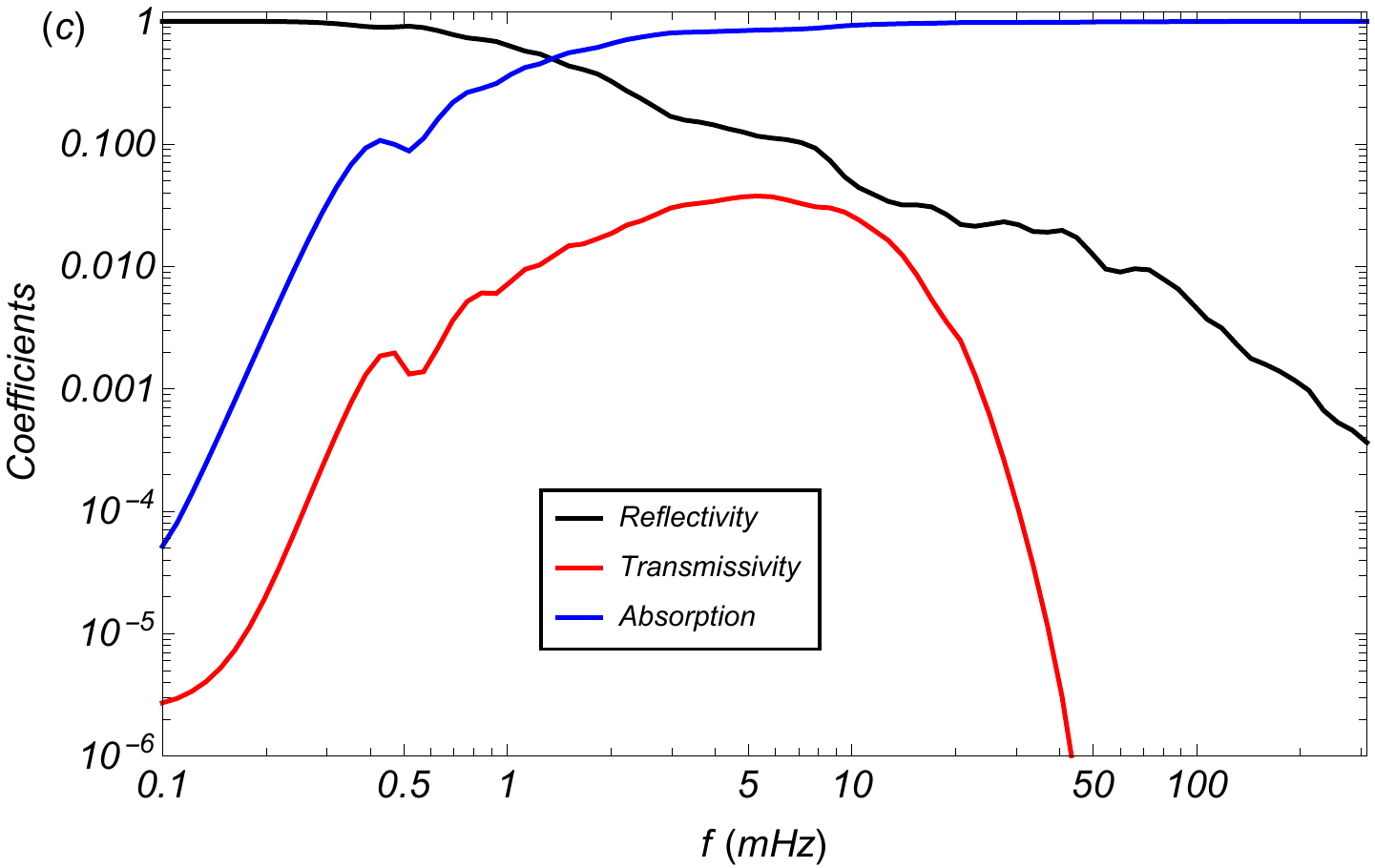}{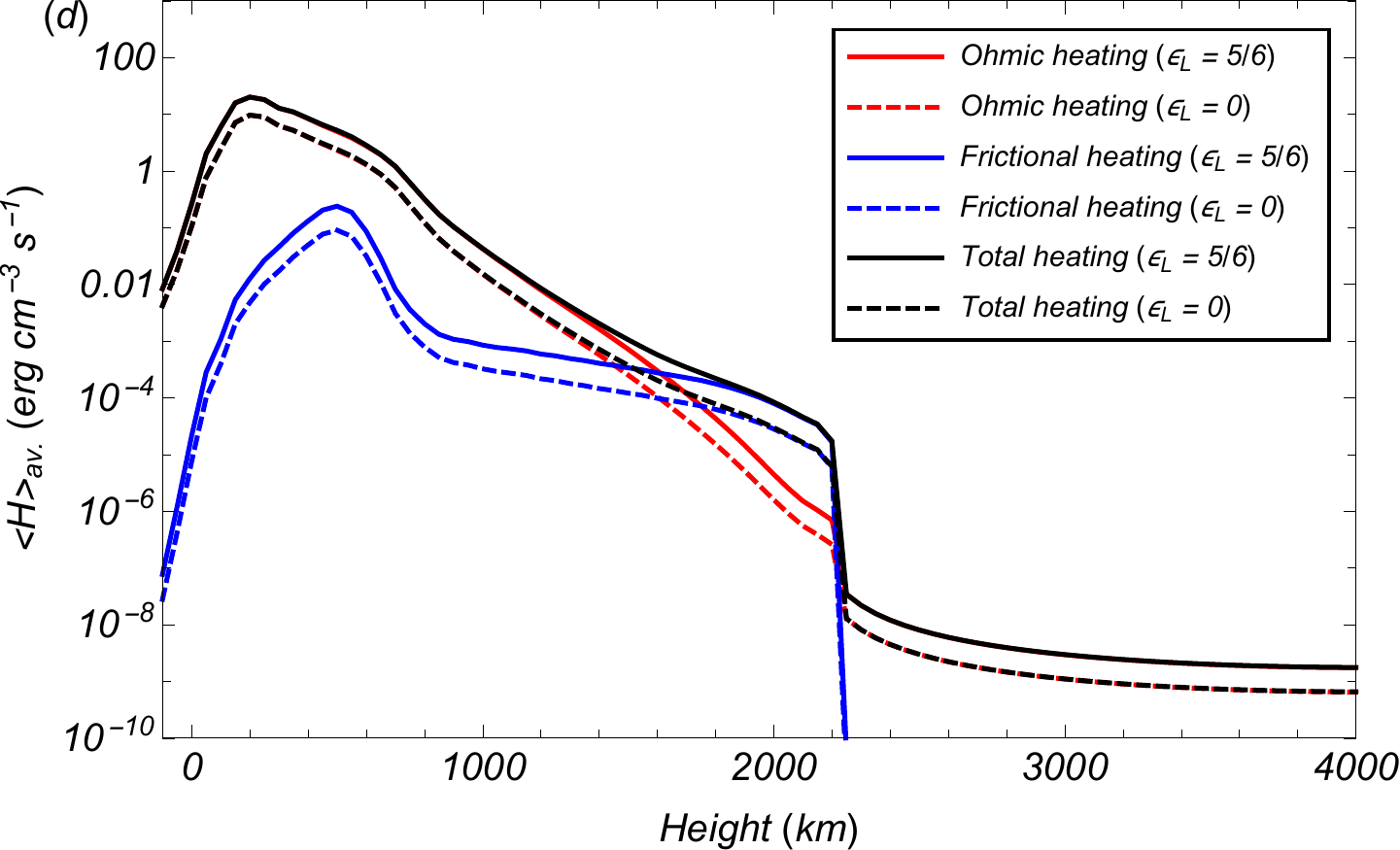}
\caption{(a) Horizontally-averaged upward, downward, and net wave energy fluxes as functions of height above the photosphere.  (b) Horizontally-averaged incident and reflected fluxes at the photospheric boundary and transmitted flux at the coronal boundary as functions of the wave frequency. (c) Coefficients of wave energy reflectivity, transmissivity, and absorption as functions of the wave frequency. (d)  Horizontally-averaged Ohmic, frictional, and total wave heating rates as functions of height above the photosphere. Results with $B_{\rm ph} = 1$~kG and for the cases with $\epsilon_L = 5/6$ and $\epsilon_L = 0$. The peak frequency of the incoming flux is $f_p = 1.59$~mHz.
\label{fig:fluxes}}
\end{figure*}

Here we turn to the study of the wave energy flux.  Figure~\ref{fig:fluxes}(a) displays the horizontally-averaged upward, downward, and net wave energy fluxes as functions of height above the photosphere. We see that the net flux, i.e., the actual energy that propagates upwards, decreases with height by several orders of magnitude. The net flux that is able to reach the corona is $\sim 1.5\times 10^5$~erg~cm$^{-2}$~s$^{-1}$, which corresponds to about 1\% of the injected flux at the photosphere. Two different mechanisms are behind this dramatic decrease of the energy flux with height, namely reflection and dissipation. Figure~\ref{fig:fluxes}(b) helps us to understand how the two processes oppose upward energy transmission.

 Figure~\ref{fig:fluxes}(b) shows the horizontally-averaged incoming (injected) and reflected fluxes at the photospheric boundary and the transmitted flux at the coronal boundary as functions of the frequency. For frequencies in the lower part of the spectrum, i.e., for $f \lesssim f_p$, the reflected flux roughly equals the incoming flux, meaning that the energy stored in those low frequencies returns back to the photosphere via reflection.  The comparison of the height-dependent upward and downward fluxes plotted in Figure~\ref{fig:fluxes}(a) reveals that most of the reflection  takes place in the middle and upper chromosphere, between 1,000~km above the photosphere and the transition region. In the lower chromosphere reflection is less important, as evidenced by the fact that the upward flux is much larger than the downward flux, while  in the coronal part of the domain reflection is virtually zero.

Returning to Figure~\ref{fig:fluxes}(b), we see that as the frequency increases, the reflected flux decreases. In turn, the transmitted flux first increases, until it reaches a maximum and eventually decreases again. The initial increase of the transmitted flux is because the wavelengths become shorter and shorter as the frequency increases. As explained in \citet{soler2017}, when the wavelengths become comparable to or smaller than the gravitational scale height, the waves can propagate with less and less reflection \citep[see also, e.g.,][]{musielak1995}. In an ideal, dissipation-less medium, the transmitted flux would monotonically increase with the frequency.  However, the chromosphere is a dissipative medium and dissipation becomes relevant for high frequencies.  Dissipation is very efficient when the frequency is in the upper part of the spectrum, i.e., for $f \gtrsim f_p$, and the damping of the waves  reduces the fraction of energy that reaches the corona. This explains the presence of a maximum in the transmitted flux and why  it later decreases rapidly and is effectively zero for the highest frequencies in the spectrum. Contrary to reflection, dissipation predominantly works in the lower chromosphere, where Ohm's diffusion is most efficient.

Another feature seen in Figure~\ref{fig:fluxes}(a) is that the  upward and net fluxes first increase sightly with height in the very low chromosphere, until $\sim 200$~km above the photosphere, before they start to decrease. The reason for this counter-intuitive  behavior is that the fluxes plotted in Figure~\ref{fig:fluxes}(a) are horizontally averaged. We recall that the average injected flux at the photosphere is $10^7$~erg~cm$^{-2}$~s$^{-1}$ but, since the photospheric filling factor is very small, the actual flux within the flux tube is larger, namely $\sim 10^9$~erg~cm$^{-2}$~s$^{-1}$. In the first 200~km above the photosphere, the  upward flux within the flux tube remains roughly constant because the accumulated effect of  reflection and dissipation is not  significant yet.  However,  the filling factor becomes larger because of the flux tube expansion. As a consequence of this, the horizontally averaged flux increases until reflection and dissipation are efficient enough to counterbalance the increase of the filling factor with height.

Although only a small fraction of the injected energy at the photosphere is transported to the corona, the transmitted energy may still be significant for the energy balance in the coronal plasma. \citet{Withbroe1977} indicate that the total energy loss in the quiet-Sun corona is $\sim 3 \times 10^5$~erg~cm$^{-2}$~s$^{-1}$, which is only twice the value of the Alfv\'enic energy flux obtained here.

\subsection{Reflectivity, transmissivity, and absorption}

In order to compare the present 2.5D results with the 1.5D results of \citet{soler2017}, we plot in Figure~\ref{fig:fluxes}(c) the coefficients of wave energy reflectivity, transmissivity, and absorption as functions of the wave frequency. These coefficients are computed by comparing the average energy fluxes at the upper and bottom boundaries of the domain (see Section~{\ref{sec:coefs}). As before, we consider the results with $B_{\rm ph} = 1$~kG, but a detailed study of the dependence of the transmission coefficient on the photospheric field strength is given later in Section~\ref{sec:fit}. Figure~\ref{fig:fluxes}(c) can be compared with Figure~4 of \citet{soler2017}. 

While the reflectivity  behaves similarly as in \citet{soler2017}, i.e., the reflectivity decreases as the frequency increases, the present results are characterized by a much larger absorption. Here, absorption starts to dominate for  frequencies much shorter than in \citet{soler2017}, so that the wave energy propagation is first dominated by reflection (for low frequencies) and later by absorption (for intermediate and high frequencies). Transmission is always residual. For the considered parameters, the maximum value of the transmissivity is $\sim 0.03$  for $f\approx 5$~mHz. As opposed to the results of  \citet{soler2017}, here there is no frequency range for which transmissivity is the largest coefficient. As discussed before, the reason for this important discrepancy is the effect of Ohm's diffusion, greatly enhanced by phase mixing. Such a relevant ingredient is absent from the computations of  \citet{soler2017}. This points out the importance of considering appropriately all the relevant dissipation mechanisms that are at work in the chromosphere.

\subsection{Heating rate}
\label{sec:heating}

The injected wave energy that is neither reflected back to the photosphere nor transmitted to the corona is dissipated in the chromosphere. The dissipated wave energy acts as a source of heating for the plasma. Figure~\ref{fig:fluxes}(d) shows the horizontally-averaged heating rate as function of height. In that figure, we display the total heating rate as well as the heating rates associated to Ohmic diffusion alone and ion-neutral friction alone. We obtain that Ohmic heating dominates throughout the chromosphere except at a relatively narrow layer just below the transition region, where frictional heating is more important. The location of this layer dominated by frictional heating is consistent with the occurrence of the maximum values of the ion-neutral drift (see Figure~\ref{fig:cutsvfn}(a)).  In turn, the largest values of the Ohmic heating rate are found at low heights where the current density perturbation is largest (see Figure~\ref{fig:cutscurrent}(a)).

The result that Ohmic diffusion is the predominant heating mechanism even at large heights in the chromosphere  somehow contradicts the estimations of its efficiency based on 1.5D models, which predicted that the importance of Ohmic diffusion would be confined to low heights \citep[see][]{soler2015scales}. In this regard, the heating rates found here at low and medium heights are somewhat larger than those obtained in the 1.5D numerical simulations of \citet[][see their Figure~6]{arber2016}. The probable reason for this difference is that damping due to Ohmic diffusion is greatly enhanced by phase mixing in our 2.5D computations compared to the less efficient damping found in 1.5D results. In fact, Figure~6 of \citet{arber2016} shows that, in their case, Ohmic heating is only dominant at heights below $\sim 500$~km, while here this dissipation mechanism remains the predominant one up to $\sim$~1,500~km. Conversely, the frictional heating rates obtained in the present 2.5D model are similar to those computed in the 1.5D case, as it can be seen by comparing Figure~\ref{fig:fluxes}(d) with Figure~8 of \citet{soler2017}.

 To determine whether the dissipated wave energy can be important for the plasma energy balance, the computed heating rate needs to be compared with the rate of energy loss due to radiation. Classic estimations of chromospheric radiative losses \citep[see][]{Withbroe1977} are $10^{-1}$~erg~cm$^{-3}$~s$^{-1}$ in the low chromosphere and $10^{-2}$--$10^{-3}$~erg~cm$^{-3}$~s$^{-1}$  in the middle and high chromosphere. The heating rates obtained here are compatible with those energy requirements, and even larger values than those required are obtained in the lower chromosphere. Dissipation of Alfv\'en waves, predominantly by Ohmic diffusion, is very efficient in the lower chromosphere, which results in large heating rates.

\subsection{Results with $\epsilon_L=0$}

Up to now, we have shown results obtained when the low-frequency exponent in the spectral weighting function is $\epsilon_L=5/6$. Here we discuss how the results are modified when we consider $\epsilon_L=0$. In Figure~\ref{fig:fluxes} we have  overplotted the results obtained for $\epsilon_L=0$. 

The overall behavior of the upward, downward, and net fluxes  is the same as for $\epsilon_L=5/6$ (see Figure~\ref{fig:fluxes}(a)), but now smaller fluxes are obtained. In this case, the transmitted flux to the corona is $\sim 6\times 10^4$~erg~cm$^{-2}$~s$^{-1}$, about half the value obtained when $\epsilon_L=5/6$. Although the injected flux at the photosphere is the same for both values of $\epsilon_L$, a smaller energy transmission to the corona is obtained when $\epsilon_L=0$. The reason for this result is that a larger fraction of the incoming flux is stored in low frequencies when $\epsilon_L=0$, as it can be seen in Figure~\ref{fig:fluxes}(b). Since those low frequencies are mostly reflected back to the photosphere, the total energy flux that is able to reach the corona is lower when $\epsilon_L=0$.

Regarding the coefficents of energy transmission, reflection, and absorption (Figure~\ref{fig:fluxes}(c)), we recall that these coefficients are intrinsic properties of the background, but are independent of the value of the injected flux and the form of the spectral weighting function. Hence, the coefficients are independent of the value of $\epsilon_L$.

Concerning the heating rates, again the overall behavior for $\epsilon_L=0$ is the same as that for $\epsilon_L=5/6$. In Figure~\ref{fig:fluxes}(d) we have also overplotted the results with $\epsilon_L=0$. As in the case of the energy flux, smaller values of the heating rate are obtained for $\epsilon_L=0$, but the differences are not significant. In both cases, the heating rates remain within the same order of magnitude. The reason is the same as before: a larger fraction of the injected energy is reflected back to the photosphere when $\epsilon_L=0$, leaving less energy to be dissipated in the chromosphere. However, as heating is mainly caused by the dissipation of high frequencies, here the impact of the value of $\epsilon_L$ is not very important.

In summary, we conclude that the form of the spectral weighting function at low frequencies, and so the amount of energy that is injected at those low frequencies, has some impact on the energy transmission to the corona because the energy of those low frequencies is mostly reflected.  Conversely,  the low-frequency exponent has a minor effect on the chromospheric heating rates, since heating is mainly caused by high frequencies.

\section{Empirical fits of the transmissivity}
\label{sec:fit}

As a practical application, we include some empirical fits of the wave energy transmissivity  that could be useful for coronal models. The results discussed so far were obtained for the case that the photospheric field strength is 1~kG. It is also interesting to generalize our results to other values of the photospheric field strength.

Figure~\ref{fig:fits}(a) shows the coefficient of wave energy transmissivity as a function of $f$ (in logaritmic scale), for four different values of $B_{\rm ph}$, namely 100~G, 500~G, 1~kG, and 2~kG. In all cases, the coronal field strength is 10~G. We find that the maximum of the transmissivity grows and is shifted towards higher frequencies when $B_{\rm ph}$ increases. The overall behavior that the transmissivity grows when the photospheric field strength increases agrees with the results obtained in \citet{soler2017} for the 1.5D case. However, even in the case with the strongest photospheric field strength,  the injected wave energy that is transmitted to the corona is a very small fraction. Hence, despite the impact of the photospheric field strength on the shape and amplitude of the transmissivity, the rest of results and the main conclusions remain qualitatively similar to those discussed in Section~\ref{sec:results} for the case of $B_{\rm ph} = 1$~kG. 

By visual inspection of Figure~\ref{fig:fits}(a), we notice that the dependence of the transmissivity on $f$ can be well approximated analytically by a skewed log-normal distribution, namely
\begin{eqnarray}
\mathcal{T}(f) &\approx& a_0 \frac{1}{\sqrt{2\pi \sigma^2}}\exp\left[ -\frac{\left(\log_{10} f - \mu\right)^2}{2\sigma^2} \right] \nonumber \\
&&\times \left[ 1 + {\rm erf}\left( \frac{\alpha}{\sqrt{2}} \frac{\log_{10} f - \mu}{\sigma} \right) \right], \label{eq:dist}
\end{eqnarray}
where $\rm erf$ is the error function, while $a_0$, $\mu$, $\sigma$, and $\alpha$ are the amplitude, location, scale, and shape parameters, respectively. We have overplotted in Figure~\ref{fig:fits}(a) the best fits obtained by using the analytic formula of Equation~(\ref{eq:dist}) and adjusting the values of the parameters $a_0$, $\mu$, $\sigma$, and $\alpha$. A very good agreement is found with the numerical results.  The $R^2$ coefficients of these fits for $B_{\rm ph}=$~100~G, 500~G, 1~kG, and 2~kG are 0.995953, 0.996036, 0.998229, and 0.998082, respectively.

\begin{figure*}
\plottwo{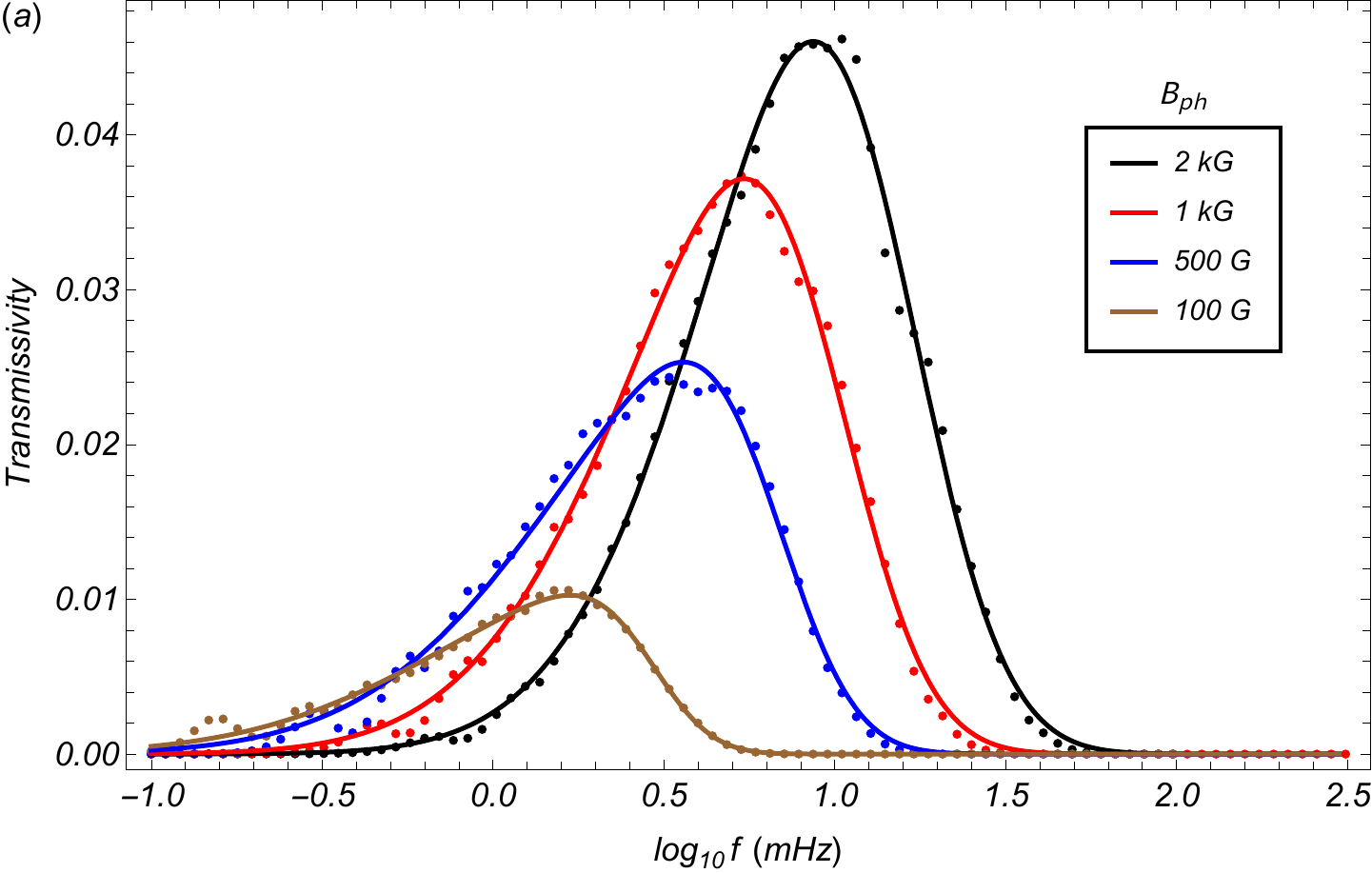}{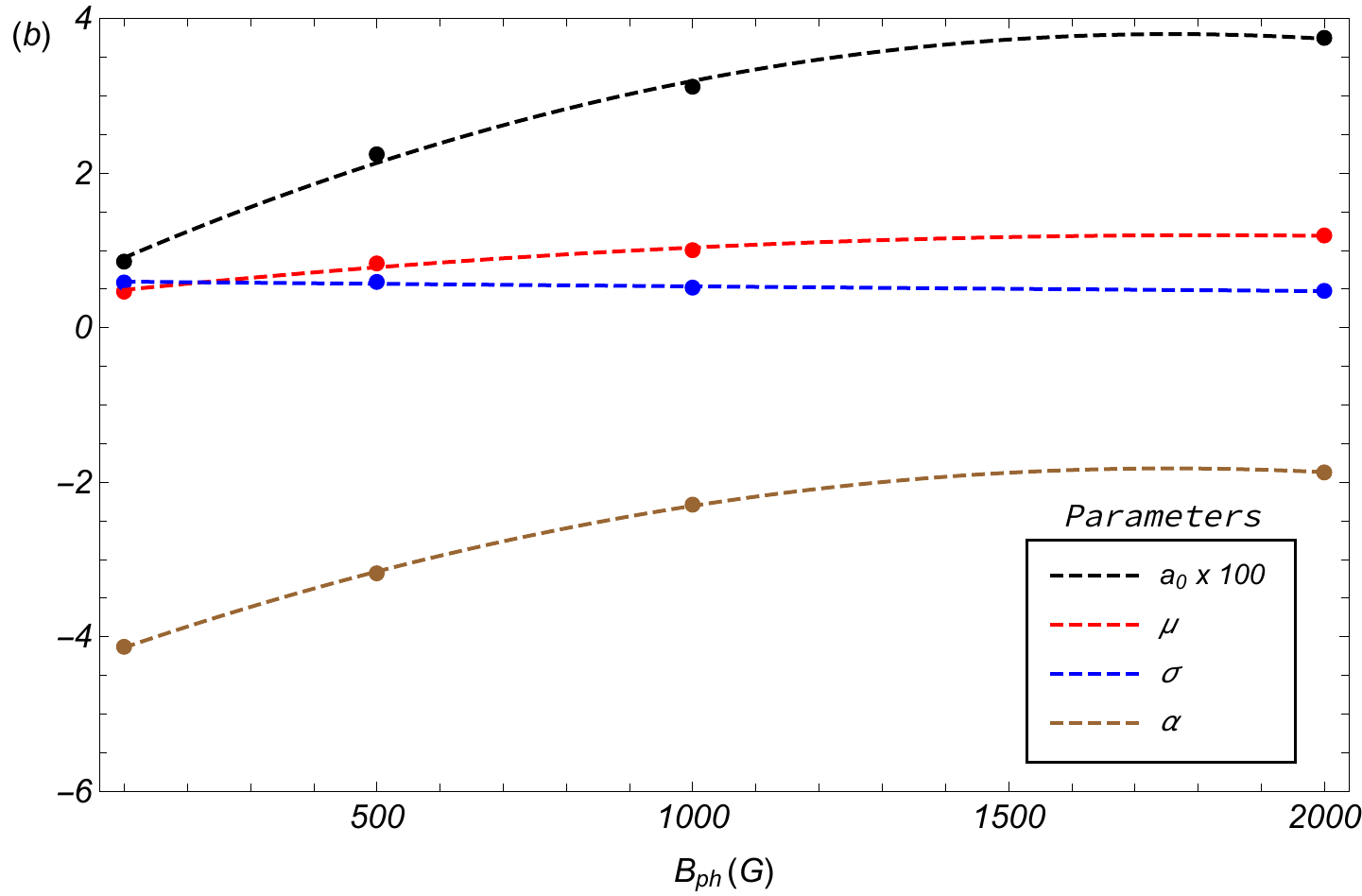}
\caption{(a) Alfv\'en wave energy transmissivity as function of $\log_{10} f$ for $B_{\rm ph}=$~100~G, 500~G, 1~kG, and 2~kG. The points are the numerical results, whereas the solid lines are the empirical fits based on Equation~(\ref{eq:dist}).  (b) Best fits of the parameters $a_0$, $\mu$, $\sigma$, and $\alpha$ in Equation~(\ref{eq:dist}) as functions of $B_{\rm ph}$. The points are the numerical results for $B_{\rm ph}=$~100~G, 500~G, 1~kG, and 2~kG, whereas the dashed lines are the results of adjusting to each parameter the  parabolic function $c_0 + c_1 B_{\rm ph} + c_2 B_{\rm ph}^2 $, where   $c_0$, $c_1$, and $c_2$ are constants   given in Table~\ref{tab:fits}. 
\label{fig:fits}}
\end{figure*}

Figure~\ref{fig:fits}(b) displays the values of $a_0$, $\mu$, $\sigma$, and $\alpha$ corresponding to the best fits as functions of $B_{\rm ph}$. Further fits have been done by adjusting each of these parameters to a parabolic function in the photospheric magnetic field strength, namely $c_0 + c_1 B_{\rm ph} + c_2 B_{\rm ph}^2 $, where $c_0$, $c_1$, and $c_2$ are constants whose specific values are given in Table~\ref{tab:fits}. 

Thus, Equation~(\ref{eq:dist}) together with the fitted parameters in Table~\ref{tab:fits} can be used to approximate the wave energy transmission coefficient for a range of photospheric field strengths between 100~G and 2~kG. This approximation  avoids the necessity of computing a full numerical solution of the waves in the photosphere and chromosphere, so it can be  used as a lower boundary condition for coronal-only models. We note that the transmissivity is independent of the  photospheric wave spectrum, which should be assumed independently in order to compute the transmitted energy.

\begin{deluxetable*}{lcccc}[!t]
\tablecaption{Parabolic fit coefficients of the parameters $a_0$, $\mu$, $\sigma$, and $\alpha$ in Equation~(\ref{eq:dist}) as functions of the photospheric magnetic field strength, $B_{\rm ph}$. The rightmost column denotes the $R^2$ coefficient of the corresponding parabolic fit. \label{tab:fits}}
\tablecolumns{5}
\tablewidth{0pt}
\tablehead{
\colhead{Parameter} &
\colhead{$c_0$} &
\colhead{$c_1$} & \colhead{$c_2$} & \colhead{$R^2$}
}
\startdata
$a_0 \times 100$ & $0.543043$  & $0.00369942$  & $-1.05127\times 10^{-6}$ & $0.995629$ \\
$\mu$ & $0.401902$ & $0.000874326$ & $-2.4108\times 10^{-7}$ & $0.985080$ \\
$\sigma$ & $0.600716$ & $-0.000071398$  & $3.4683\times 10^{-9}$ & $0.894329$  \\
$\alpha$ & $-4.43062$ & $0.00296635$ & $-8.43723\times 10^{-7}$ & $0.999716$ \\
\enddata
\tablecomments{Each parameter has been fitted to a parabolic function $c_0 + c_1 B_{\rm ph} + c_2 B_{\rm ph}^2 $, where the photospheric magnetic field strength, $B_{\rm ph}$, is given in G.  We note that these fits assume that the wave frequency, $f$,  in the skewed log-normal distribution (Equation~(\ref{eq:dist})) is given in mHz.}
\end{deluxetable*}

\section{Concluding remarks}
\label{sec:conc}

In this work we have studied the energy transport and dissipation associated with  torsional Alfv\'en waves that propagate through the solar atmosphere from the photosphere towards the corona. We have performed an extension and improvement of the previous work of \citet{soler2017} by incorporating new relevant ingredients, namely the consideration of a 2.5D  model and the presence of the term due to Ohm's magnetic diffusion in the induction equation. Both additions turned out to have important impacts on the results.

On the one hand, the 2.5D model used here allowed us to fully solve the radial dependence of the wave perturbations, which was not possible in the 1.5D models used in the previous literature. As a consequence of that, the effect of phase mixing is present here. The magnetic shear generated by phase mixing across the flux tube produces large current density perturbations, which are efficiently dissipated by Ohmic diffusion in the low and mid chromosphere. This results in a stronger damping of the Alfv\'en waves in those regions compared to estimations based on 1.5D models.

On the other hand, and in connection to the above comment, the presence of  Ohm's magnetic diffusion dramatically reduces the net upward energy flux of the waves. Ohmic diffusion is the dominant damping mechanism in most  of the chromosphere, while ion-neutral collisions are only dominant in the higher chromosphere. This fact does not imply that the role of partial ionization is not important, on the contrary. We recall that the Ohmic diffusivity is greatly enhanced in partially ionized plasmas because of electron-neutral collisions (see again Figure~\ref{fig:background}(c)), so that the very efficient role of  Ohmic diffusion in the chromosphere is a direct consequence of the plasma being partially ionized.

Another effect that decreases the  upward energy flux is reflection, whose role is especially important in the upper chromosphere and transition region. Because of reflection, counter-propagating waves co-exist in the chromosphere. It is known that the interaction of counter-propagating Alfv\'en waves can lead to plasma turbulence, which has been proposed as another important heating mechanism in the solar atmosphere \citep[see, e.g.,][]{vanballegooijen2011}. The role of turbulence has not been incorporated here but could be explored in future works.

The combined effects of dissipation and reflection cause that only about 1\% of the wave energy flux driven at the photosphere is able to reach coronal heights. Although small, the transmitted energy flux may still represent a significant energy input for the coronal plasma when compared to the total coronal energy loss in quiet-Sun conditions \citep[see][]{Withbroe1977}. We have provided an empirical fit of the transmission coefficient that could be used to incorporate in coronal models the energy flux of photospherically driven Alfv\'en waves. The use of these empirical formulae would avoid the need of considering the very narrow chromosphere with a sufficiently high resolution  to actually resolve the wave transmission.  

We have computed the heating rates associated with the dissipation of wave energy, which seem to be compatible with the chromospheric energy requirements. However, since we have restricted ourselves to the linear regime, we cannot actually compute the plasma thermalization associated  with the deposition of this heating in the form of internal energy. To do so, we should use the full, nonlinear equations and should also consider the effects of radiation losses and thermal conduction. This cannot be done in the steady-state assumption used here and should be done, necessarily, using time-dependent simulations.

Owing to the lack of detailed observational information, our choice for the photospheric wave driver relies on two  assumptions, namely the value of the average incoming energy flux and the form of the spectral weighting function. Regarding the incoming flux, we have used $10^7$~erg~cm$^{-2}$~s$^{-1}$, which is the value typically assumed in the all the recent literature \citep[see, e.g.,][]{depontieu2001,goodman2011,tu2013,arber2016} and is based on some numerical estimations \citep[see][]{Ulmschneider2000}. Obviously, increasing/decreasing this value would result in a larger/smaller energy transmission to the corona. Concerning the spectral weighting function, changing the weight that low/high frequencies have in the spectrum would affect the fraction of the total wave  energy that is reflected/dissipated. That, in turn, would also modify the transmitted flux. Future high-resolution observations are needed to shed some light on the nature of the photospheric wave drivers.

An approximation in our model is that there is no reflection at the upper coronal boundary. The condition that there are no incoming waves from the corona is necessary because our model only includes the lower corona up to 4,000 km above the photosphere. We do not incorporate information about the coronal structure at larger heights, and so we ignore the possible reflections that may occur there. In the context of plasma heating by turbulence, Alfvén wave reflection in the corona has been studied by, e.g., \citet{matthaeus1999,cranmer2005,zank2018}. In relation with the purpose of this paper, i.e., to study wave energy propagation from the photosphere to the corona through the chromosphere, the results show that the reflection of waves driven at the photosphere is important in the high chromosphere and transition region but, in our calculations, reflection does not play a relevant role above the transition region since only about 1\% of the driven energy flux is able to reach those heights (see again  Figure~\ref{fig:fluxes}(a)). Even if all the transmitted flux was reflected back to the chromosphere, the results discussed here would not be modified significantly.

Finally, we should note that the static model used here represents an idealization of the actual atmosphere. In reality, the chromosphere is a very dynamic medium. The study of  Alfv\'en waves in more realistic, time-varying models is a challenge because of the tremendous complexity that represents separating the wave activity  from the  dynamic evolution of the background \citep[see, e.g.,][]{khomenko2018}. Early attempts to understand wave behavior in dynamic plasmas have been undertaken \citep[see][]{ballester2018}. However, it is necessary to keep improving the models and to approach them to reality in order to advance our understanding of the role of the waves in the solar atmosphere dynamics.

\acknowledgments
We acknowledge the support from grant AYA2017-85465-P (MINECO/AEI/FEDER, UE). RS acknowledges the `Ministerio de Econom\'ia, Industria y Competitividad' and the `Conselleria d'Innovaci\'o, Recerca i Turisme del Govern Balear (Pla de ci\`encia, tecnologia, innovaci\'o i emprenedoria 2013-2017)' for the `Ram\'on y Cajal' grant RYC-2014-14970.

\bibliographystyle{aasjournal} 
\bibliography{refs}

\end{document}